\documentclass[%
 aip,
 amsmath,amssymb,
preprint,%
]{revtex4-1}

\usepackage{graphicx}
\usepackage{dcolumn}
\usepackage{bm}

\usepackage[utf8]{inputenc}
\usepackage[T1]{fontenc}
\usepackage{mathptmx}
\usepackage{etoolbox}

\newcommand{\pd}[2]{\frac{\partial #1}{\partial #2}}
\newcommand{\ave}[1]{\left\langle #1 \right\rangle}

\makeatletter
\def\@email#1#2{%
 \endgroup
 \patchcmd{\titleblock@produce}
  {\frontmatter@RRAPformat}
  {\frontmatter@RRAPformat{\produce@RRAP{*#1\href{mailto:#2}{#2}}}\frontmatter@RRAPformat}
  {}{}
}%
\makeatother
\begin{document}


\title[Spectral transport budgets of turbulent heat fluxes in PCT]{Spectral analysis on transport budgets of turbulent heat fluxes in plane Couette turbulence}
\author{T. Kawata}
 \affiliation{ 
Keio University, Hiyoshi 3-14-1, Yokohama, 223-8522 Kanagawa, Japan
}
 \email{kawata@keio.jp}
 
\author{T. Tsukahara}%
 \affiliation{ 
Tokyo University of Science, Yamazaki 2641, Noda, 278-8510 Chiba, Japan
}%

\date{\today}

\begin{abstract}
In recent years, scale-by-scale energy transport in wall turbulence has been intensively studied, and the complex spatial and interscale transfer of turbulent energy has been investigated. As the enhancement of heat transfer is one of the most important aspects of turbulence from an engineering perspective, it is also important to study how turbulent heat fluxes are transported in space and in scale by nonlinear multi-scale interactions in wall turbulence as well as turbulent energy. In the present study, the spectral transport budgets of turbulent heat fluxes are investigated based on direct numerical simulation data of a turbulent plane Couette flow with a passive scalar heat transfer. The transport budgets of spanwise spectra of temperature fluctuation and velocity-temperature correlations are investigated in detail in comparison to those of the corresponding Reynolds stress spectra. The similarity and difference between those scale-by-scale transports are discussed, with a particular focus on the roles of interscale transport and spatial turbulent diffusion. As a result, it is found that the spectral transport of the temperature-related statistics is quite similar to those of the Reynolds stresses, and in particular inverse interscale transfer is commonly observed throughout the channel in both transport of the Reynolds shear stress and wall-normal turbulent heat flux.\end{abstract}

\maketitle


\section{\label{sec:intro} Introduction}
Analysing the budgets of the Reynolds stress transport equation is a useful tool to investigate turbulence transport phenomena, as it shows quantitatively where and how much turbulence is produced from the mean flow, redistributed among different velocity components by the effect of pressure, dissipated into heat by viscosity, etc. A particularly interesting feature is that the budget includes spatial transport (or diffusion) terms, which do not lead to total energy gain/loss of turbulent energy across the flow field but represent spatial transports by different effects, namely, advection by the mean flow and diffusion by turbulence, pressure fluctuation and viscosity. Similarly to the Reynolds stress transport equation, one can derive the transport equations of passive-scalar fluxes, and one of the most engineering-relevant features of turbulence to enhance the spatial transport of heat and mass may be expressed by the turbulent diffusion terms in these transport equations. 

The budget analysis of the Reynolds stress (or turbulent kinetic energy) transport equation has recently been extended to scale-by-scale budget analysis, where the transport equation is decomposed into large- and small-scale parts~(e.g., Refs.~\onlinecite{kawata_2018,wang_2020,hwang_2021,kawata_2021,chin_2021}) or the spectral transport equation~\cite{mizuno_2016,lee_2019,wang_2021} is derived. Similar analysis of the energy transport at each scale is also possible based on the transport equations of two-point correlation quantities, such as the second-order structure function~(e.g., Refs.~\onlinecite{cimarelli_2013,cimarelli_2016,gatti_2021}). The main feature of such scale-by-scale budget analysis is that the turbulent diffusion term is split into the spatial diffusion at each scale and the interscale transport at each spatial location, which allows us to separately investigate the spatial and interscale transfer effects by the nonlinear scale interaction of turbulence. 

Such scale-by-scale budget analysis of the turbulent energy transport has been used to study the complex energy transport in wall turbulence, and the behaviour of the interscale energy transport term was particularly focused on in relation to the dynamics of coherent structures in the inner and outer layers. As these inner and outer structures have clearly different spanwise length scales~\cite{smits_2011}, many of the earlier studies on the scale-by-scale budget analysis were based on spanwise length scales (e.g., Refs.~\onlinecite{cho_2018,kawata_2018,chin_2021,hwang_2021}), and it has been shown that the interscale energy transport terms exhibit both forward (from larger to smaller scales) and backward (from smaller to larger scales) energy transfers in the near-wall region~\cite{mizuno_2016,cho_2018,kawata_2018,hamba_2018,gatti_2021,hwang_2021}, unlike in homogeneous isotropic turbulence, where the energy transfer is basically from larger to smaller scales. 

The scale-by-scale analysis of the passive-scalar fluxes can also be performed by the almost same manner as the analysis of the Reynolds stress transport as their transport equations are similar. Such scale-by-scale analysis of passive-scalar transfer allows us to investigate the interscale and spatial transport of the passive scalar caused by nonlinear interactions between different scales in the velocity fields. However, despite numerous studies reported on the wall turbulence with a passive scalar field (e.g., Refs.~\onlinecite{kawamura_1998,kawamura_1999,hane_2006,tsukahara_2007t,antonia_2009,abe_2007,avila_2018,avila_2019,avila_2021a,avila_2021b}) in the last few decades, the scale-by-scale transport of the passive scalar in wall turbulence has not been explored yet. 

In the present study, we investigate the transport equation budgets of the temperature-related spectra based on the direct numerical simulation (DNS) data of wall turbulence with a passive-scalar heat transfer, with a particular focus on the roles of the interscale and spatial transport at each scale. Turbulent plane Couette flow is chosen as the test case, because very-large-scale structures filling up the entire channel appear at relatively low Reynolds numbers~\cite{komminaho_1996,kitoh_2005,tsukahara_2006,kitoh_2008,avsarkisov_2014,orlandi_2015,lee_2018}, and therefore, scale separation between the inner and outer structures is relatively clearer than in other canonical wall-bounded flow configurations at the same Reynolds number. The constant-temperature-difference condition is adopted as the thermal boundary condition on the walls so that the mean velocity and temperature profiles are similar to each other. Then, fluctuating velocity and temperature fields are compared by focusing on the similarity/difference in the interscale and spatial energy transport caused by scale interactions. 
In our previous paper~\cite{kawata_2022a}, the discussion was mainly based on streamwise Fourier-mode analysis, and the budget analysis based on spanwise scale was not focused on. This paper investigates the spectral transport of the temperature fluctuation and the turbulent heat fluxes based on the spanwise Fourier mode. In particular, our interest lies in the transport of the wall-normal turbulent heat flux in comparison with that of the Reynolds shear stress, where inverse interscale transfer is observed throughout the channel.

\section{\label{sec:setup} DNS Dataset and Spectral Budget Analysis}

\subsection{DNS Dataset of a Turbulent Plane Couette Flow with Temperature Transport}

In the present study, the budget analysis of the spectral transport equations is performed with a DNS dataset of a turbulent plane Couette flow with passive-scalar temperature transport obtained in our previous work~\cite{kawata_2022a}. Here, we briefly summarise the computational conditions. The geometrical configuration of the plane Couette flow is such that a shear flow is driven by the stationary bottom wall and the top wall translating with a constant speed $U_\mathrm{w}$, which are separated by a distance $h$. The $x$-, $y$-, and $z$-axes are taken in the streamwise, wall-normal, and spanwise directions, and the origin of the coordinates is placed on the stationary bottom wall; the bottom and top walls are located at $y=0$ and $y=h$, respectively. The temperatures of the walls are uniform and constant in time, and the temperature difference is $\Delta T=T_\mathrm{t}-T_\mathrm{b}$ ($T_\mathrm{t}$ and $T_\mathrm{b}$ are the temperature of the top and the bottom walls, respectively).  

\begin{table}
    \caption{Computational conditions: streamwise and spanwise length of computational domain, $L_x^\ast$ and $L_z^\ast$; number of grid points in each direction, $N_x$, $N_y$, and $N_z$; spatial resolutions, $\Delta x$, $\Delta y$, and $\Delta z$; friction Reynolds number $Re_\tau$. $\Delta y_\mathrm{max}$ and $\Delta y_\mathrm{min}$ are the maximum and minimum value of the wall-normal spatial resolution, respectively, and the superscript $^+$ for the spatial resolutions stands for the values scaled by viscous length $\nu^\ast/u_\tau^\ast$. The outer Reynolds number and the Prandtl number are $Re_\mathrm{w}=8600$ and $Pr=0.71$, respectively. }
    \label{tab:num}\vspace{0.3\baselineskip}
    \begin{ruledtabular}
    \begin{tabular}{ccccc}
        $(L_x^\ast,L_z^\ast)$ & $(N_x,N_y,N_z)$ & $(\Delta x^+,\Delta z^+)$ & $(\Delta y^+_\mathrm{max},\Delta y^+_\mathrm{min})$ & $Re_\tau$\\
        $(24h,12.8h)$ & $(512,96,512)$ & $(11.8,6.31)$ & $(6.16,0.268)$ & $126.1$
    \end{tabular}
    \end{ruledtabular}
\end{table}

The governing equations of fluid flow are the continuity and Navier-Stokes equations for incompressible flow that are non-dimensionalised by $h$ and $U_\mathbf{w}$, and the convection-diffusion equation of passive-scalar temperature that was by $\Delta T$, $h$, and $U_\mathbf{w}$ was solved for the fluctuating temperature field. A non-slip boundary condition was applied to the fluid velocity on the wall, and as described above the temperature on the walls are uniform and constant in time with the temperature difference $\Delta T$. It should be noted here that both the velocity and temperature fields are driven by a uniform and constant velocity or temperature difference between the top and bottom walls, and therefore the boundary conditions on the walls for the velocity and temperature fields are similar when scaled by the velocity difference $U_\mathrm{w}$ and the temperature difference $\Delta T$, respectively. For the streamwise and spanwise directions, periodic boundary conditions were applied. The Reynolds number $Re=U_\mathrm{w} h/\nu$ and Prandtl number $Pr=\alpha/\nu$ are $Re_\mathbf{w}=8600$ and $Pr=0.71$, where $\nu$ and $\alpha$ are the kinematic viscosity and the thermal diffusivity of the fluid, respectively, and other details of the DNS, such as the domain size and spatial resolution, are listed in Table~\ref{tab:num}. 

\begin{figure}[t!]
    \centering
    \includegraphics[width=0.6\hsize]{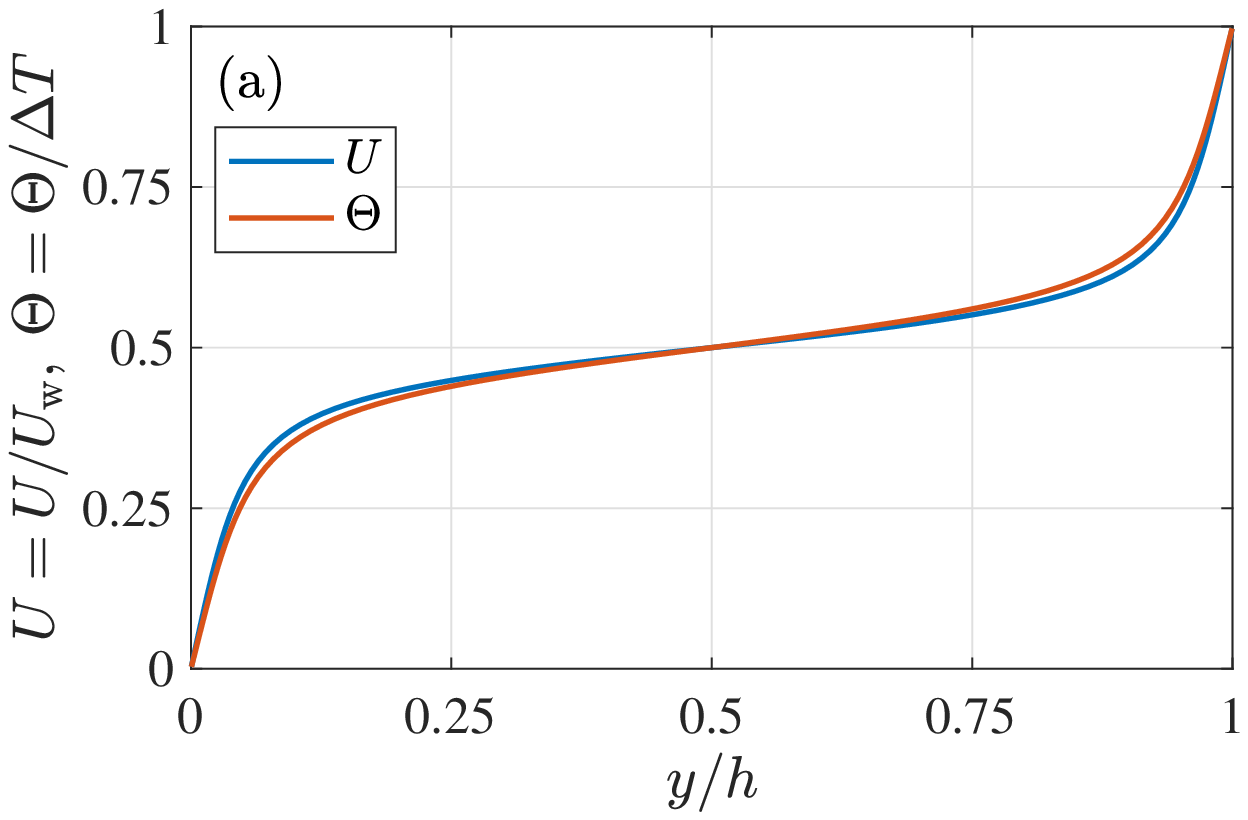}
    \includegraphics[width=0.6\hsize]{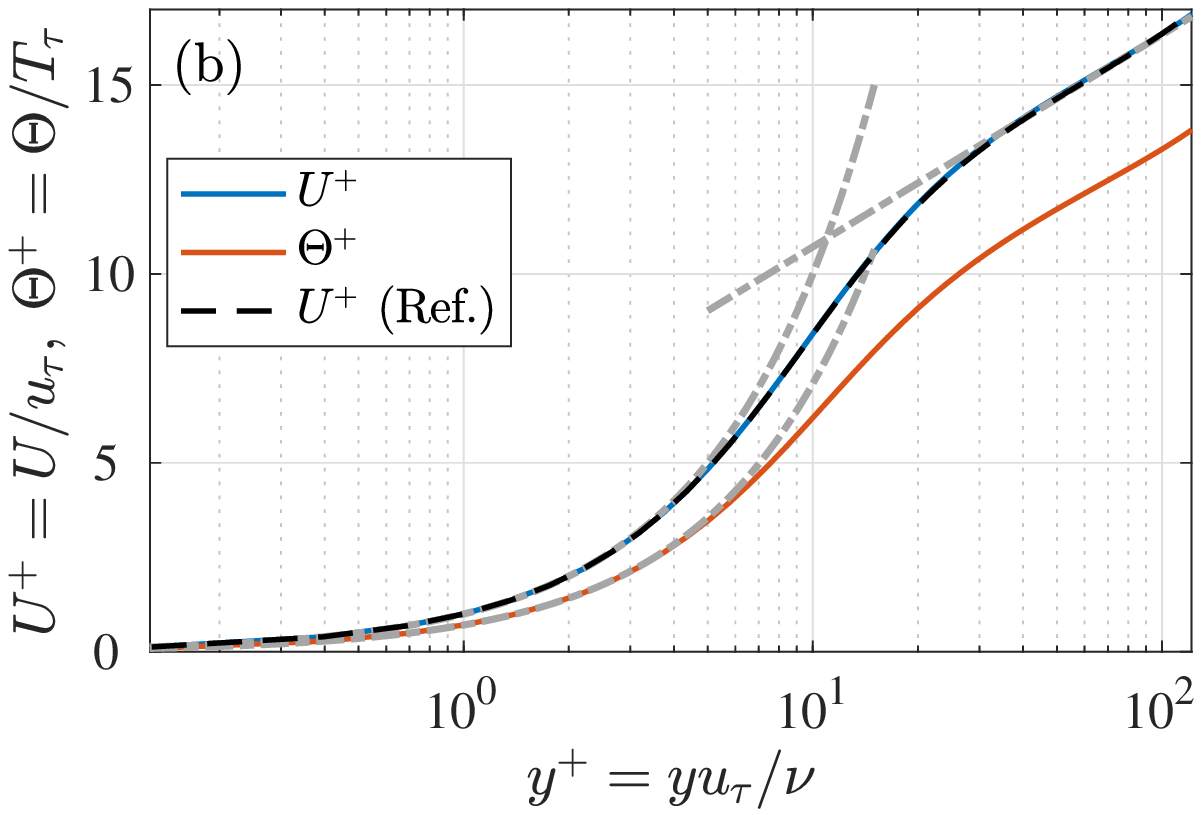}
    \caption{Profiles of mean streamwise velocity $U$ and the mean temperature $\varTheta$ in (top) outer and (bottom) inner scaling. In the bottom panel, the grey chained lines indicate $U^+=y^+$ and $U^+=2.44 \ln y^+ +5.1$ for the mean velocity profile and $\varTheta^+=Pr y^+$ for the mean temperature profile. The black dashed line in the panel~(b) presents reference data~\cite{kawata_2021} obtained with an extremely large computational domain~$(L_x, L_z)=(96.0h, 12.8h)$. }
    \label{fig:meanUT}
\end{figure}

We denote the velocity components in the $x$-, $y$-, and $z$-directions as $U+u$, $v$, $w$, respectively. Here $U$ is the mean streamwise velocity and the lower-case letter represents the fluctuation around the mean value of each velocity component (note here that the mean wall-normal and spanwise velocities are zero). The temperature is defined as the temperature difference between the fluid and the bottom wall and denoted as $\varTheta + \theta$: here $\varTheta$ and $\theta$ are the mean temperature and the temperature fluctuation, respectively, and $\varTheta=0$ and $\Delta T$ at the bottom and top walls, respectively. In the following, $\ave{}$~represents averaged quantities obtained by averaging in $x$- and $z$-directions and in time. They are also averaged between the lower and upper half of the channel because of the symmetric flow configuration. 

\begin{figure}
    \centering
    \includegraphics[width=0.65\hsize]{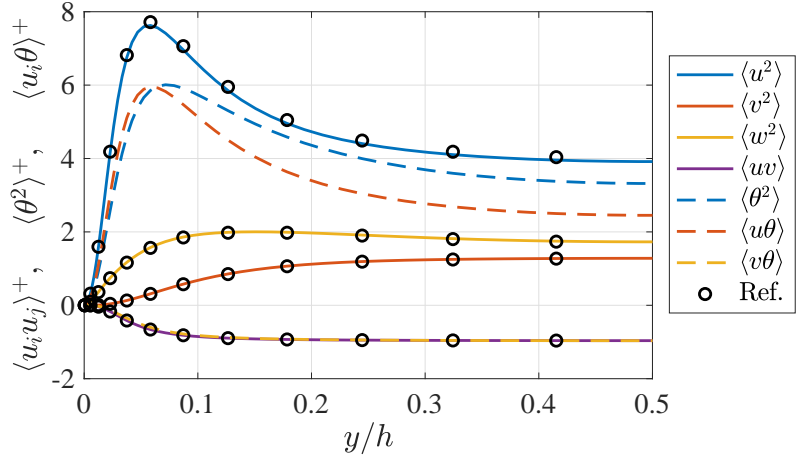}
    \caption{Profiles of (solid lines) the Reynolds stresses $\ave{u_i u_j}$ and (dashed lines) the temperature-related turbulent statistics $\ave{\theta^2}$ and $\ave{u_i \theta}$. The values are scaled based on the inner units, $u_\tau$ and/or $T_\tau$, and only $0 \ll y \ll 0.5$ is shown. The black circles present reference data for the Reynolds stresses~\cite{kawata_2021}. }
    \label{fig:rs}
\end{figure}

Figures~\ref{fig:meanUT} and \ref{fig:rs} show the profiles of the mean velocity and temperature and those of the second-order statistics, such as velocity and temperature variances and their cross correlations. Here, $T_\tau$ used in Fig.~\ref{fig:meanUT}(b) is the friction temperature defined as $T_\tau=Q_\mathrm{w} / \rho c_p u_\tau$, where $Q_\mathrm{w}$ and $c_p$ are the mean heat flux on the wall and the specific heat at constant pressure, respectively, and $u_\tau$ is the friction velocity defined as $u_\tau=\sqrt{\tau_\mathrm{w}/\rho}$ ($\tau_\mathrm{w}$ and $\rho$ are the wall shear stress and the fluid density). As shown in Fig.~\ref{fig:meanUT}(a), the profile of the mean temperature $\varTheta$ is anti-symmetric similarly to the mean velocity profile due to the similar boundary conditions for the velocity and temperature fields, and it is also seen in Fig.~\ref{fig:meanUT}(b) that the mean temperature gradient $\mathrm{d}\varTheta/\mathrm{d}y$ in the vicinity of the wall is markedly smaller than the mean streamwise velocity gradient $\mathrm{d}U/\mathrm{d}y$. This difference can be attributed to the fact that $Pr=0.71<1$, which makes the effect of molecular diffusion more significant in the temperature field than in the velocity field. 

Figure~\ref{fig:rs} presents the profiles of the Reynolds stresses $\ave{u^2}$, $\ave{v^2}$, $\ave{w^2}$, and $\ave{uv}$, and the temperature-related statistics: the temperature fluctuation $\ave{\theta^2}$ and the velocity-temperature correlations $\ave{u\theta}$ and $\ave{v\theta}$. As shown here, the temperature fluctuation $\ave{\theta^2}$ and the streamwise-velocity-temperature correlation $\ave{u\theta}$ have similar profiles to the streamwise velocity fluctuation $\ave{u^2}$, which indicates a certain similarity between the fluctuations of the streamwise velocity $u$ and of the temperature $\theta$. This is attributable to the fact that the profiles of the mean streamwise velocity $U$ and mean temperature $\varTheta$ are similar. It is also shown here that the profiles of the Reynolds shear stress $\ave{uv}$ and turbulent heat flux $\ave{v\theta}$ are almost on top on each other, which indicates strong similarity between the turbulent momentum and the heat transfers. 

In the DNS of turbulent plane Couette flow, one needs to use a considerably large computational domain to exclude the dependency of the obtained DNS results on the computational domain size due to the very-large-scale structure~\cite{komminaho_1996}. The effect of the domain size was carefully addressed in our previous work~\cite{kawata_2021}, and it was confirmed that the domain size used for obtaining the present DNS dataset, $(L_x,L_z)=(24.0 h, 12.8 h)$, was large enough to exclude the domain-size effect, as demonstrated in Figs.~\ref{fig:meanUT} and \ref{fig:rs} by the good agreement between the present computational results and those obtained with a much larger computational domain~\cite{kawata_2021}. 

\subsection{Spectral Analysis of the Reynolds-Stress Transport and Turbulent Heat Transfers} \label{sec:analysis}

In the present study, the spectral transports of the temperature fluctuation $\ave{\theta^2}$ and the turbulent heat fluxes $\ave{u_i \theta}$ are analysed based on spanwise Fourier mode analysis. The transport equations of the temperature-related spectra are derived similarly to those for the Reynolds stress spectra. Here, the derivation of the transport equations of the Reynolds stress spectra is briefly reviewed, and the spectral transport equations for the temperature-related statistics are introduced by similar means. 

\subsubsection{Transport of Reynolds stresses}

The transport of the Reynolds stresses $\ave{u_i u_j}$ is expressed as
\begin{subequations} \label{eq:rste}
\begin{equation}
\left(\pd{}{t}+U_k\pd{}{x_k}\right) \ave{u_i u_j} = P_{ij}-\varepsilon_{ij}+\Phi_{ij}+D^\nu_{ij}+D^t_{ij}, 
\end{equation}
where the terms on the right-hand side are the production ($P_{ij}$), viscous dissipation ($\varepsilon_{ij}$), pressure-gradient work ($\Phi_{ij}$), viscous diffusion ($D^\nu_{ij}$), and turbulent diffusion ($D^t_{ij}$) terms, which are defined as
\begin{align}
P_{ij}&=-\ave{u_i u_k}\pd{U_j}{x_k}-\ave{u_j u_k}\pd{U_i}{x_k}, \quad
\varepsilon_{ij} = 2 \nu \ave{\pd{u_i}{x_k}\pd{u_j}{x_k}},\\
\Phi_{ij} &= -\ave{\pd{p}{x_i}u_j}-\ave{\pd{p}{x_j}u_i}, \quad
D^\nu_{ij} = \nu \pd{^2 \ave{u_i u_j}}{x_k^2}, \quad 
D^t_{ij} = -\pd{\ave{u_i u_j u_k}}{x_k}.
\end{align}
\end{subequations}
The pressure term $\Phi_{ij}$ may be decomposed into the pressure-strain correlation $\Pi_{ij}$ and the pressure diffusion $D^p_{ij}$:
\begin{subequations} \label{eq:Pi_ij}
\begin{align}
    \Phi_{ij} = \Pi_{ij} + D^p_{ij}, \quad
    \mathrm{where} \quad \Pi_{ij} &= \ave{p \left( \pd{u_i}{x_j} + \pd{u_j}{x_i}\right)}, \\
                   D^p_{ij} &= - \pd{}{x_k} \left( \ave{u_i p} \delta_{jk} + \ave{u_j p} \delta_{ik}\right).
\end{align} 
\end{subequations}
$\delta_{ij}$ is the Kronecker's delta.

Then, we consider the decomposition of fluctuating velocities and temperature into their large- and small-scale parts 
\begin{subequations}\label{eq:dec}
\begin{align} 
    u_i = u_i^L + u_i^S, \quad \theta=\theta^L+\theta^S,
\end{align}
at a cutoff spanwise wavenumber $k_{z,c}$ based on the spanwise Fourier modes. With such a decomposition the cross correlation between the large- and small-scale parts is zero for any combination of quantities:
\begin{align}
    \ave{u_i^L u_j^S} = \ave{u_j^L u_i^S} = 0,  \quad
    \ave{\theta^L \theta^S} = \ave{u_i^L \theta^S} = \ave{u_i^S \theta^L} = 0. 
\end{align}
\end{subequations}
Hence, the Reynolds stresses $\ave{u_i u_j}$ are decomposed simply into their large- and small-scale parts as
\begin{align} \label{eq:rsdec}
    \ave{u_i u_j} = \ave{u_i^L u_j^L} + \ave{u_i^S u_j^S},
\end{align}
and by a similar manner to derive the ``full'' Reynolds stress equation (\ref{eq:rste}) one obtains the transport equations of the large- and small-scale parts of the Reynolds stresses (the details of the derivation are found in Refs.~\onlinecite{kawata_2019,kawata_2021}) as
\begin{subequations} \label{eq:rstels}
\begin{align}
\left( \pd{}{t}+U_k\pd{}{x_k} \right) \ave{u_i^L u_j^L}= P^L_{ij}-\varepsilon^L_{ij}+\Phi^L_{ij}+D^{\nu,L}_{ij}+D^{t,L}_{ij} -Tr^z_{ij}, \\
\left( \pd{}{t}+U_k\pd{}{x_k} \right) \ave{u_i^S u_j^S} = P^S_{ij}-\varepsilon^S_{ij}+\Phi^S_{ij}+D^{\nu,S}_{ij}+D^{t,S}_{ij}+Tr^z_{ij}. 
\end{align}
\end{subequations}
The terms on the right-hand sides of Eqs.~(\ref{eq:rstels}a, b) are the large- and small-scale parts of their counterpart in Eq.~(\ref{eq:rste}), except for $Tr_{ij}$. In particular, the first four terms on the right-hand side of each equation (i.e., the production, viscous dissipation, pressure-gradinet work, and viscous diffusion) are simply decomposed into their large- and small-scale parts similarly to the Reynolds stress decomposition in Eq.~(\ref{eq:rsdec}): these terms in Eq.~(\ref{eq:rstels}a) are defined as those in Eq.~(\ref{eq:rste}b-e) with $u_i$ and $u_j$ replaced by $u_i^L$ and $u_j^L$ (the corresponding terms in Eq.~(\ref{eq:rstels}b) are similarly defined with $u_i^S$ and $u_j^S$). 

On the other hand, the turbulent diffusion is decomposed as  
\begin{align}
D^{t,L}=-\pd{\ave{u_i u_j u_k}^L}{x_k}, \quad 
D^{t,S}=-\pd{\ave{u_i u_j u_k}^S}{x_k},
\end{align}
where $\ave{u_i u_j u_k}^L$ and $\ave{u_i u_j u_k}^S$ are, respectively, the large- and small-scale part of the triple velocity correlation $\ave{u_i u_j u_k}$ defined as
\begin{subequations} \label{eq:dectri}
\begin{align}
\ave{u_i u_j u_k}^L=\ave{u_i^L u_j^L u_k^L}+\ave{u_i^L u_j^L u_k^S}+\ave{u_i^S u_j^L u_k^S}+\ave{u_i^L u_j^S u_k^S},\\
\ave{u_i u_j u_k}^S=\ave{u_i^S u_j^S u_k^S}+\ave{u_i^S u_j^S u_k^L}+\ave{u_i^L u_j^S u_k^L}+\ave{u_i^S u_j^L u_k^L},
\end{align}
\end{subequations}
and the last term in Eqs.~(\ref{eq:rstels}a, b), $Tr^z_{ij}$, is defiend as 
\begin{align} \label{eq:Trij}
Tr^z_{ij} = 
-\ave{u_i^S u_k^S \pd{u_j^L}{x_k}}-\ave{u_j^S u_k^S \pd{u_i^L}{x_k}} 
+\ave{u_i^L u_k^L \pd{u_j^S}{x_k}}+\ave{u_j^L u_k^L \pd{u_i^S}{x_k}}.
\end{align}
It is observed from Eqs.(\ref{eq:rstels}a, b) that this term appears in the transport equations of both $\ave{u_i^L u_j^L}$ and $\ave{u_i^S u_j^S}$ with different signs, which means that $Tr^z_{ij}$ represents the energy flux between the large- and small-scale velocity fields across the cutoff wavenumber $k_{z,c}$. 

It is worth pointing out here that only the turbulent diffusion terms ($D^{t,L}_{ij}$ and $D^{t,S}_{ij}$) and the interscale energy flux term ($Tr^z_{ij}$) include both the large- and small-scale parts of the fluctuating velocities, whereas the other terms in the transport equations consist of either the large- or small-scale part only. This is because the turbulent transport and the interscale energy flux terms are the third-order moments of the fluctuating velocities (and the velocity gradients), while the other terms are second-order moments. These terms including both the large- and small-scale parts of the fluctuating velocity can be interpreted as the energy transport effects by interactions between different scales: the turbulent transport ($D^{t,L}_{ij}$ and $D^{t,S}_{ij}$) indicates the energy transport in physical space caused by scale interactions, while the interscale energy flux ($Tr^z_{ij}$) is the energy transport between different scales. 

As the large-scale part of the Reynolds stresses $\ave{u_i^L u_j^L}$ and the spanwise one-dimensional spectra of the Reynolds stresses are related as 
\begin{align} \label{eq:eij_uiuj}
    E^z_{ij} = \pd{\ave{u_i^L u^L_j}}{k_{z,c}} \left(=-\pd{\ave{u_i^S u_j^S}}{k_{z,c}} \right),
\end{align}
the transport equations of the Reynolds-stress spectra can be derived by differentiating both sides of Eq.~(\ref{eq:rstels}a) with respect to $k_{z,c}$ as
\begin{align} \label{eq:eijte}
\left( \pd{}{t}+U_k\pd{}{x_k} \right) E^z_{ij} =
pr^z_{ij}-\xi^z_{ij}+\phi^z_{ij}+d^{\nu,z}_{ij}+d^{t,z}_{ij}+tr^z_{ij},
\end{align}
where the terms on the right-hand side are the $k_{z,c}$-derivatives of the counter parts in Eq.~(\ref{eq:rstels}a). The first five terms represent the spectra of the corresponding terms of Eq.(~\ref{eq:rste}a): the spectra of the production ($pr^z_{ij}$), viscous dissipation ($\xi^z_{ij}$), pressure term ($\phi^z_{ij}$), viscous diffusion ($d^{\nu,z}_{ij}$) and turbulent diffusion ($d^{t,z}_{ij}$) terms. The interscale transport term~$tr^z_{ij}$ represents the energy gain/loss by the interscale flux $Tr_{ij}$ at each wavenumebr $k_z$. 
According to Eq.~(\ref{eq:Pi_ij}), the pressure-gradient-work spectrum $\phi_{ij}$ can also be decomposed into the spectra of the pressure-strain energy redistribution term $\Pi_{ij}$ and the pressure diffusion term $D^p_{ij}$ as 
\begin{align} \label{eq:pi_ij}
    \phi^z_{ij} = \pi^z_{ij} + d^{p,z}_{ij}.
\end{align}

The turbulent spatial transport $d^{t,z}_{ij}$ and the interscale transport $tr^z_{ij}$, which are of particular interest in the present study, are defined as
\begin{subequations}
\begin{align}
d^{z,t}_{ij}=-\pd{E^z_{ijk}}{x_k}, \quad\quad tr^z_{ij} = - \pd{Tr^z_{ij}}{k_{z,c}}.
\end{align}
Here, $E^z_{ijk}$ is the spectra of triple velocity correlations $\ave{u_i u_j u_k}$ defined as
\begin{align}
E^z_{ijk} = \pd{\ave{u_i u_j u_k}^L}{k_{z,c}} = -\pd{\ave{u_i u_j u_k}^S}{k_{z,c}}.
\end{align}
\end{subequations}
As the physical meaning of $\ave{u_i u_j u_k}$ is the spatial transport flux of the Reynolds stress $\ave{u_i u_j}$ in the $x_k$-direction caused by the velocity fluctuation $u_k$, their spectra $E^z_{ijk}$ represent the spatial flux of the Reynolds stresses $\ave{u_i u_j}$ in the $x_k$-direction at each spanwise wavenumber.  

\subsubsection{Transport of temperature fluctuation and velocity-temperature correlations}

Now, we derive the spectral transport equations of the temperature-related turbulence statistics. The transport equation of the temperature fluctuation $\ave{\theta^2}$ is written similarly to the Reynolds stress equation (\ref{eq:rste}) as
\begin{subequations}\label{eq:ttte}
\begin{equation}
\left(\pd{}{t}+U_k\pd{}{x_k}\right) \ave{\theta^2} = P_{\theta \theta}-\varepsilon_{\theta \theta}+D^\nu_{\theta \theta}+D^t_{\theta \theta} ,
\end{equation}
with the production ($P_{\theta\theta}$), viscous dissipation ($\varepsilon_{\theta\theta}$), viscous diffusion ($D^\nu_{\theta\theta}$), and turbulent transport ($D^t_{\theta\theta}$) terms defined as
\begin{align}
P_{\theta \theta} = - 2\ave{\theta u_k}\pd{\varTheta}{x_k}, \quad
\varepsilon_{\theta\theta} = 2 \alpha \ave{\pd{\theta}{x_k} \pd{\theta}{x_k}}, \quad
D^\nu_{ij} = \alpha \pd{^2 \ave{\theta^2}}{x_k^2}, \quad 
D^t_{\theta\theta} = -\pd{\ave{\theta^2 u_k}}{x_k}.
\end{align}
\end{subequations}
The transport equation of the turbulent heat fluxes $\ave{u_i \theta}$ is also similarly obtained as
\begin{subequations}\label{eq:tvte}
\begin{equation}
\left(\pd{}{t}+U_k\pd{}{x_k}\right) \ave{u_i \theta} = P_{i \theta}-\varepsilon_{i \theta}+\Phi_{i\theta}+D^\nu_{i \theta}+D^t_{i \theta},
\end{equation}
with the production ($P_{i \theta}$), viscous dissipation ($\varepsilon_{i \theta}$), pressure-gradient-temperature correlations ($\Phi_{i \theta}$), viscous diffusion ($D^\nu_{i \theta}$), and turbulent diffusion ($D^t_{i \theta}$) terms defined as
\begin{align}
P_{i\theta}&=-\ave{u_k \theta} \pd{U_i}{x_k} - \ave{u_i u_k}\pd{\varTheta}{x_k}, \quad
\varepsilon_{i\theta} = \left( \nu + \alpha \right) \ave{\pd{u_i}{x_k} \pd{\theta}{x_k}}, \quad
\Phi_{i\theta} = -\ave{\pd{p}{x_i} \theta}, \\
D^\nu_{i\theta} &=  \pd{}{x_k}
\left(\nu \ave{\pd{u_i}{x_k}\theta} + \alpha \ave{\pd{\theta}{x_k}u_i}
\right), \quad \quad
D^t_{i\theta} = -\pd{\ave{u_i\theta u_k}}{x_k}.
\end{align}
\end{subequations}

Based on the decomposition given by Eq.~(\ref{eq:dec}), the temperature variance and the turbulent heat fluxes are also split into their large- and small-scale parts, similarly to the Reynolds stress decomposition (\ref{eq:rsdec}), as
\begin{align}
    \ave{\theta^2} = \ave{\theta^L \theta^L}+\ave{\theta^S \theta^S},\quad
    \ave{u_i \theta} = \ave{u_i^L \theta^L} + \ave{u_i^S \theta^S}.
\end{align}
Therefore, the transport equations of their large- and small-scale parts can be derived similarly to Eqs.~(\ref{eq:rstels}). Then, these large- and small-scale parts of the temperature-related statistics are related to their spectra, similarly to Eq.~(\ref{eq:eij_uiuj}), as
\begin{align}
    E^z_{\theta \theta} = \pd{\ave{\theta^L \theta^L}}{k_{z,c}}, \quad
    E^z_{i \theta} = \pd{\ave{u_i^L \theta^L}}{k_{z,c}},
\end{align}
and the transport equations of these temperature-related spectra can also be obtained by differentiating the equations of $\ave{\theta^L \theta^L}$ and $\ave{u_i^L \theta^L}$. With such procedures one obtains the temperature-variance spectrum equation as 
\begin{align} \label{eq:ettte}
\left( \pd{}{t}+U_k\pd{}{x_k} \right) E^z_{\theta \theta}  = pr^z_{\theta\theta} - \xi^z_{\theta\theta} + d^{\nu,z}_{\theta\theta} + d^{t,z}_{\theta\theta} + tr^z_{\theta\theta},
\end{align}
with the terms on the right-hand side representing the spectra of the counterparts in Eq.~(\ref{eq:ttte}a). 
The interscale and the turbulent spatial transport terms are defined as
\begin{subequations} \label{eq:si_tt}
\begin{align}
tr^z_{\theta\theta} = - \pd{Tr^z_{\theta\theta}}{k_{z,c}}, \quad\quad d^{t,z}_{\theta\theta} = - \pd{E^z_{\theta\theta k}}{x_k}, 
\end{align}
where $Tr^z_{\theta \theta}$ and $E^z_{\theta \theta k}$ are the interscale and spectral spatial fluxes of the temperature fluctuation defined as
\begin{align} 
Tr^z_{\theta\theta} &= -2\ave{\theta^S u_k^S \pd{\theta^L}{x_k}} -2\ave{\theta^L u_k^L \pd{\theta^S}{x_k}}, \\ 
E^z_{\theta\theta k} &= \pd{\ave{\theta^2 u_k}^L}{k_{z,c}} = -\pd{\ave{\theta^2 u_k}^S}{k_{z,c}},
\end{align}
with $\ave{\theta^2 u_k}^L$ and $\ave{\theta^2 u_k}^S$ being the large- and small-scale part of the spatial flux of temperature fluctuation $\ave{\theta^2 u_k}$:
\begin{align}
\ave{\theta^2 u_k}^L &= \ave{\theta^L \theta^L u_k^L} +\ave{\theta^L \theta^L u_k^S}+ 2\ave{\theta^L \theta^S u_k^S}, \\
\ave{\theta^2 u_k}^S &= \ave{\theta^S \theta^S u_k^S} + \ave{\theta^S \theta^S u_k^L} + 2\ave{\theta^L \theta^S u_k^L}.
\end{align}
\end{subequations}

Similarly, the transport equation of the turbulent heat flux spectrum $E^z_{i\theta}$ is obtained as
\begin{align} \label{eq:eitte}
 \left( \pd{}{t}+U_k\pd{}{x_k} \right) E^z_{i \theta} = pr^z_{i\theta} - \xi^z_{i\theta} +\phi^z_{i\theta}+ d^{\nu,z}_{i\theta} + d^{t,z}_{i\theta} + tr^z_{i\theta},
\end{align}
and the terms on the right-hand side are, again, the spectra of their counterparts in Eq.~(\ref{eq:tvte}a).
The interscale and the turbulent spatial transport terms are defined in the same manner as in other spectral transport equations: 
\begin{subequations} \label{eq:si_ut}
\begin{align}
tr^z_{i\theta} = - \pd{Tr^z_{i\theta}}{k_{z,c}}, \quad\quad d^{t,z}_{i\theta} = - \pd{E^z_{i\theta k}}{x_k},
\end{align}
where the interscale flux $Tr^z_{i\theta}$ and spectral spatial flux $E^z_{i\theta k}$ are, respectively, defined as 
\begin{align}
Tr^z_{i\theta} &= -\ave{\theta^S u_k^S \pd{u_i^L}{x_k}} -\ave{u_i^S u_k^S \pd{\theta^L}{x_k}} 
 +\ave{\theta^L u_k^L \pd{u_i^S}{x_k}} +\ave{u_i^L u_k^L \pd{\theta^S}{x_k}}, \\ 
E^z_{i\theta k} &= \pd{\ave{u_i \theta u_k}^L}{k_{z,c}} = -\pd{\ave{u_i \theta u_k}^S}{k_{z,c}}.
\end{align}
Here, $\ave{u_i \theta u_k}^L$ and $\ave{u_i \theta u_k}^S$ are the large- and small-scale parts of $\ave{u_i \theta u_k}$ defined as
\begin{align}
\ave{u_i \theta u_k}^L = \ave{u_i^L \theta^L u_k^L} + \ave{u_i^L \theta^L u_k^S} + \ave{u_i^L \theta^S u_k^S} + \ave{u_i^S \theta^L u_k^S}, \\
\ave{u_i \theta u_k}^S = \ave{u_i^S \theta^S u_k^S} + \ave{u_i^S \theta^S u_k^L} + \ave{u_i^S \theta^L u_k^L} + \ave{u_i^L \theta^S u_k^L}.
\end{align}
\end{subequations}

\section{Results}\label{sec:result}

\subsection{Spectra of Temperature-Related Statistics and Their Interscale and Spatial Transports}

\begin{figure}
    \centering
    \includegraphics[width=1\hsize]{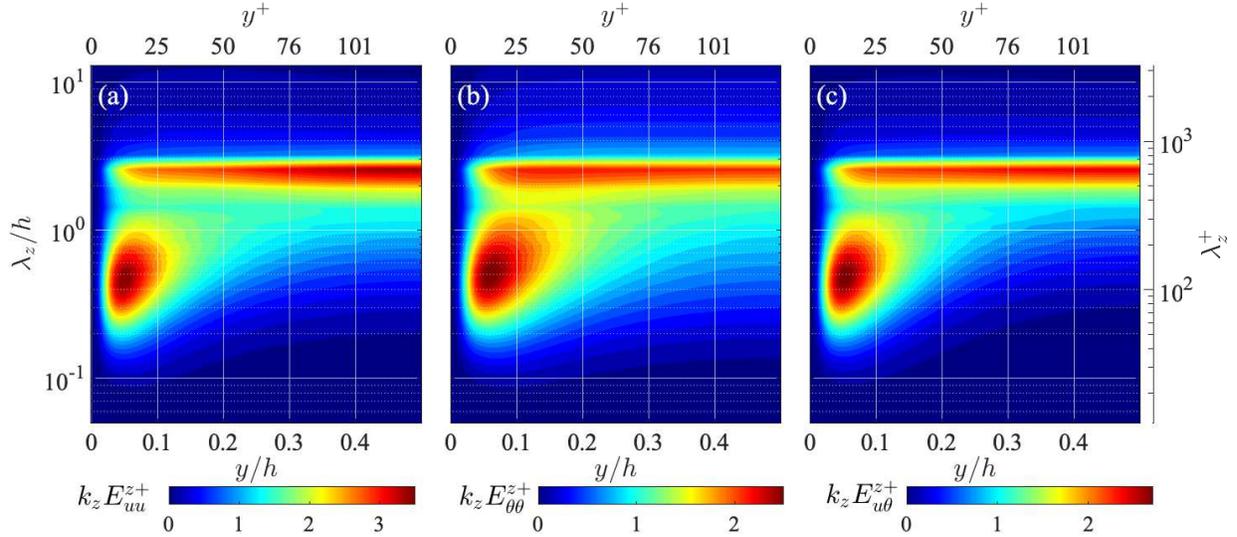}
    \caption{Space-wavelength ($y$-$\lambda_z$) diagrams of spanwise one-dimensional premultiplied spectra of (a) streamwise turbulent energy $k_z E^z_{uu}$, (b) temperature fluctuation $k_z E^z_{\theta \theta}$, and (c) velocity-temperature correlation $k_z E^z_{u \theta}$. The values are scaled by $u_\tau^2$, $T_\tau^2$, and $u_\tau T_\tau$, respectively. }
    \label{fig:ez}
\end{figure}

Figure~\ref{fig:ez} presents  the distributions of the spanwise one-dimensional spectra of the streamwise turbulent energy $E^z_{uu}$, temperature fluctuation $E^z_{\theta \theta}$, and temperature-velocity correlation $E^z_{u\theta}$ in the premultiplied form. The distribution of the streamwise turbulent energy spectrum $E^z_{uu}$ in the panel~(a) shows two significant energy concentrations, one of which is the energy peak located in a relatively small wavelength range near the wall around $(y^+,\lambda_z^+) \approx (12,100)$ and the other is the broad energy band at a large wavelength around $\lambda_z/h \approx 2.5$ covering most of the channel. These energy peaks clearly correspond to the small-scale coherent structures near the wall and the very-large-scale structure of plane Couette turbulence, respectively. As shown in this figure, the energy peaks corresponding to these coherent structures at different scales are observed to be clearly separated from each other despite the relatively small Reynolds number $Re_\tau = 126$, which is a unique feature of plane Couette turbulence \cite{kitoh_2005,tsukahara_2006,kitoh_2008}. 

The temperature-fluctuation spectrum $E^z_{\theta \theta}$ presents a similar distribution to the streamwise turbulent energy $E^z_{uu}$ as shown in Fig.~\ref{fig:ez}(b), in which the energy peaks corresponding to the near-wall and very-large-scale structures are indicated at almost the same position in the $y$-$\lambda_z$ diagram. The location of the near-wall peak is, however, at a slightly larger wavelength and further from the wall than the corresponding peak in the $E^z_{uu}$ distribution, $(y^+,\lambda_z^+) \approx (16, 134)$, which can be attributed to the more significant effect of molecular diffusion in the temperature field as $Pr<1$.  The distribution of the temperature-velocity cospectrum $E^z_{u\theta}$ also shows a similar distribution to $E^z_{uu}$, indicating similar behaviours of the fluctuating streamwise velocity and temperature. 

\begin{figure}[t!]
    \centering
    \includegraphics[width=0.7\hsize]{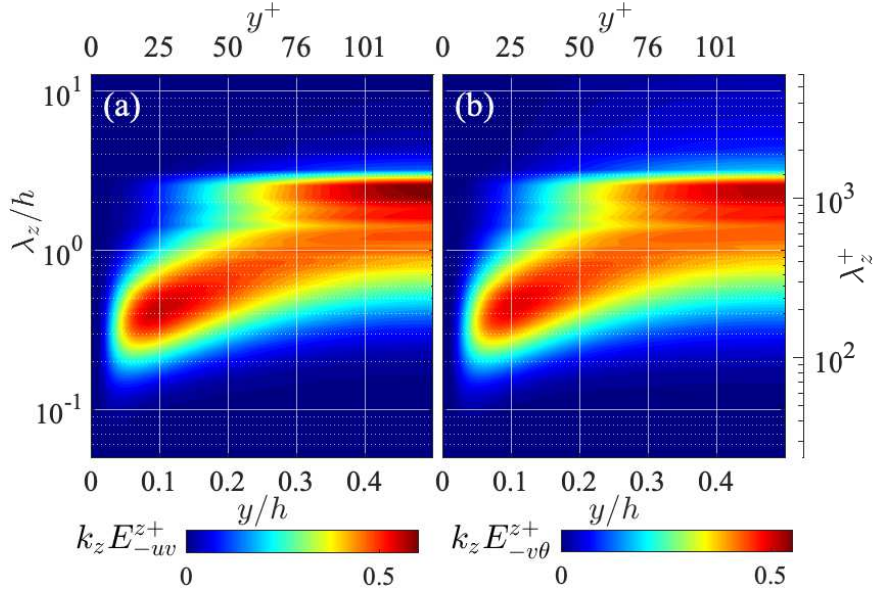}
    \caption{Spanwise one-dimensional spectra of (a) the Reynolds shear stress $E^z_{-uv}$ and (b) turbulent heat transfer $E^z_{-v \theta}$, presented in the same manner as in Fig.~\ref{fig:ez}. }
    \label{fig:uv_vtz}
\end{figure}

Figure~\ref{fig:uv_vtz} compares the spanwise one-dimensional cospectra of the Reynolds shear stress $E^z_{-uv}$ and the velocity-temperature correlation $E^z_{-v \theta}$. As the Reynolds shear stress $-\ave{uv}$ and the temperature-velocity correlation $-\ave{v \theta}$ represent the wall-normal transport of momentum and heat by turbulent fluid motions, respectively, their cospectra represent the turbulent momentum and heat transfers at each length scale. As shown in Fig.~\ref{fig:uv_vtz}(a), the distribution of the Reynolds shear stress cospectrum $E^z_{-uv}$ presents both the inner peak at small scales near the wall and the broad energy peak at large scales at the channel centre, similarly to the distribution of the streamwise turbulent energy spectrum $E^z_{uu}$. As shown in Fig.~\ref{fig:uv_vtz}(b), the distribution of the turbulent heat transfer spectrum $E^z_{-v \theta}$ is qualitatively similar to that of the momentum transfer spectrum~$E^z_{-uv}$, presenting both the inner and outer energy peaks corresponding to the coherent structures in the near-wall and channel-central regions of the channel, which indicates a certain scale-by-scale similarity between the momentum and heat transfers by turbulence. 

\subsection{Interscale Fluxes of the Reynolds Stresses and Temperature-Related Statistics}

\begin{figure}
    \centering
    \includegraphics[width=1\hsize]{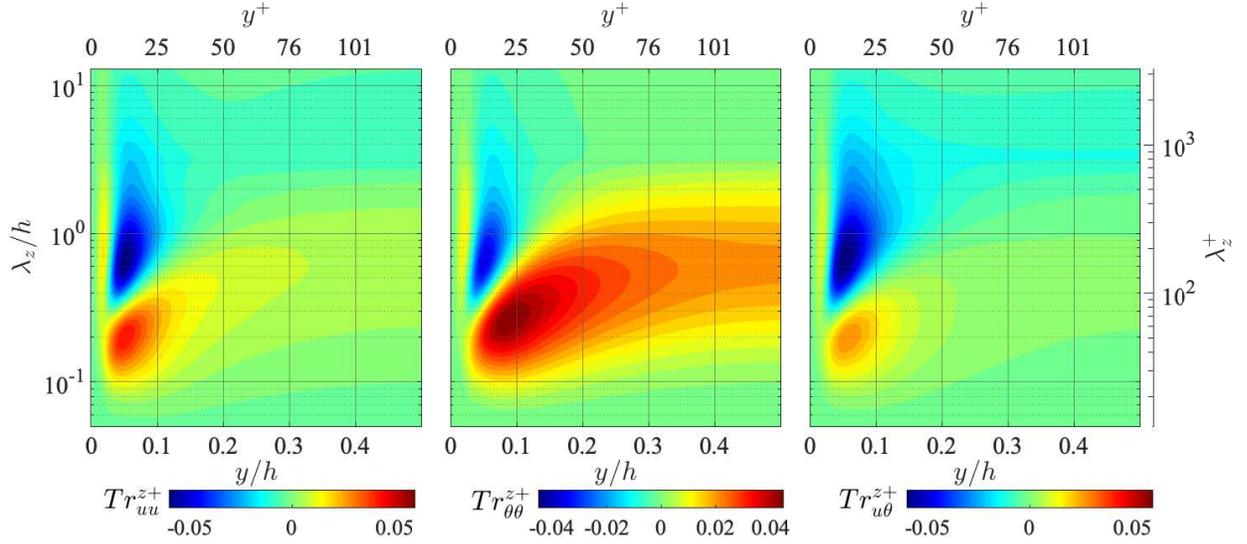}
    \caption{Space-wavelength ($y$-$\lambda_z$) diagrams of spanwise interscale fluxes of (a) the turbulent kinetic energy~$Tr^z_{uu}$, (b) temperature fluctuation~$Tr^z_{\theta\theta}$, and (c) temperature-velocity correlation $Tr^z_{u\theta}$. The values are scaled by $u_\tau^4/\nu$, $u_\tau^2$, $T_\tau^2/\nu$, and $u_\tau^3 T_\tau/\nu$, respectively. }
    \label{fig:trzuu_tt}
\end{figure}

Next, the interscale fluxes of the temperature-related spectra are investigated. Figure~\ref{fig:trzuu_tt} presents the distributions of the spanwise interscale flux of the streamwise turbulent energy~$Tr^z_{uu}$, temperature fluctuation $Tr^z_{\theta \theta}$, and temperature-velocity correlation $Tr^z_{u\theta}$. Here, as can be seen from Eq.~(\ref{eq:rstels}), the positive values indicate forward energy fluxes (i.e., from larger to smaller scales) in the spanwise length-scale direction, whereas the negative values represent the inverse (from smaller  to larger scales)  energy fluxes. As shown in Fig.~\ref{fig:trzuu_tt}(a), the turbulent energy flux $Tr^z_{uu}$ indicates mainly forward interscale energy transfers from larger to smaller $\lambda_z$ in the relatively small $\lambda_z$ range throughout the channel, but it also presents backward energy fluxes from smaller to larger $\lambda_z$ in a relatively large $\lambda_z$ range in the near-wall region. It is interesting to note that that such inverse interscale energy transfer is not observed in the streamwise length-scale direction~\cite{lee_2019,kawata_2021}, and the corresponding physical phenomenon has still not been elucidated. 

As shown in Fig.~\ref{fig:trzuu_tt}(b), the spanwise interscale flux of the temperature fluctuation $Tr^z_{\theta \theta}$ presents a qualitatively similar distribution to the streamwise turbulent energy flux $Tr^z_{uu}$, presenting a region of backward interscale energy transfers at relatively large $\lambda_z$ in the near-wall region and forward interscale energy fluxes at smaller $\lambda_z$ throughout the channel. One can, however, also observe in the distribution of $Tr^z_{\theta \theta}$ that the region of backward energy flux somewhat shrinks and instead the magnitude of the forward energy flux around the channel centre is relatively stronger than $Tr^z_{uu}$. 

Figure~\ref{fig:trzuu_tt}(c) shows the distribution of the spanwise interscale flux of the velocity-temperature correlation $Tr^z_{u\theta}$. One can observe that $Tr^z_{u\theta}$ also has a qualitatively similar distribution to those of the turbulent energy flux $Tr^z_{uu}$ and the temperature fluctuation flux $Tr^z_{\theta \theta}$, presenting both backward and forward energy flux. Furthermore, as compared to the distributions of $Tr^z_{uu}$ and $Tr^z_{\theta \theta}$, the region of backward energy flux is relatively large and the peak magnitude of the forward energy flux at small $\lambda_z$ near the wall is relatively small. 

\begin{figure}
\centering
    \includegraphics[width=0.7\hsize]{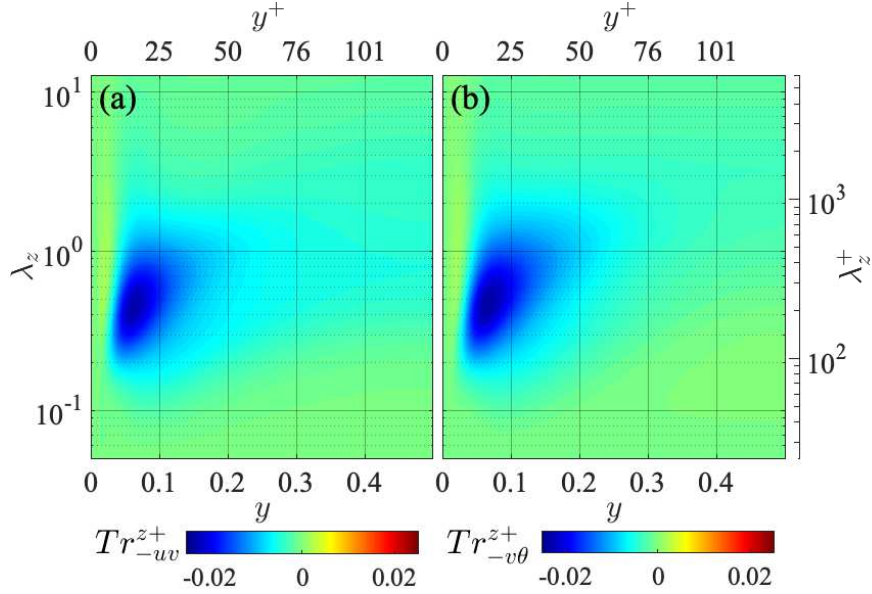}
    \caption{Spanwise interscale fluxes of (a) Reynolds shear stress~$Tr^z_{-uv}$ and (b) turbulent heat flux~$Tr^z_{-v \theta}$, presented in the same manner as in Fig.~\ref{fig:trzuu_tt}.}
    \label{fig:trzuv_vt}
\end{figure}

The spanwise interscale fluxes of the Reynolds shear stress $Tr^z_{-uv}$ and the turbulent heat flux~$Tr^z_{-v \theta}$ are also presented in Fig.~\ref{fig:trzuv_vt}. As shown in the panel~(a), the Reynolds shear stress flux $Tr^z_{-uv}$ indicates, interestingly, the backward transfers throughout the channel, as first experimentally observed by Kawata \& Alfredsson~\cite{kawata_2018}. This is in contrast to the turbulent energy transfer $Tr^z_{uu}$, which mainly indicates forward transfers throughout the channel except the near-wall region. Such inverse interscale transport of the Reynolds shear stress has also been reported in turbulent channel~\cite{gatti_2021} and boundary-layer~\cite{chin_2021} flows.

The interscale transfer of the turbulent heat flux $Tr^z_{-v \theta}$ gives a similar distribution to that of $Tr^z_{-uv}$, as presented in Fig.~\ref{fig:trzuv_vt}(b). The difference between the distributions of these spanwise interscale fluxes is that $Tr^z_{-v \theta}$ spans to slightly larger spanwise wavelengths than $Tr^z_{-uv}$. While it is still unclear what physical phenomenon is represented by such an inverse interscale transfer of $-\ave{uv}$ and $-\ave{v\theta}$, the similar distributions of $Tr^z_{-uv}$ and $Tr^z_{-v\theta}$ indicate close analogy between the transport of the Reynolds shear stress and turbulent heat flux. 
 
\subsection{Spectra of Spatial Turbulent Fluxes of the Reynolds Stresses and Temperature-Related Statistics}

\begin{figure}[t!]
     \centering
     \includegraphics[width=0.6\hsize]{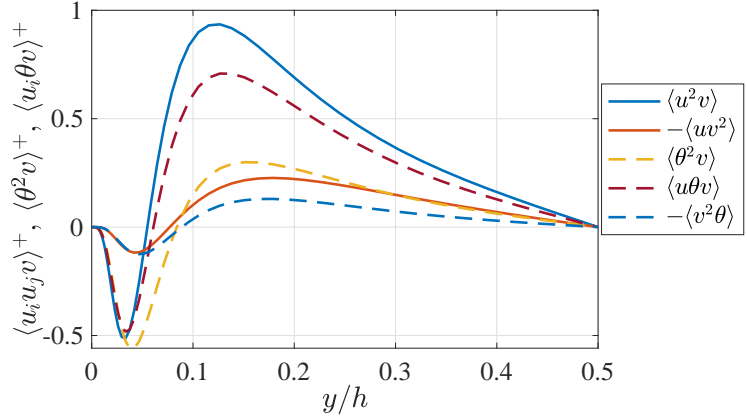}
     \caption{Profiles of the velocity triple correlations $\ave{u^2v}$ and $-\ave{uv^2}$ and the temperature-velocity triple correlations $\ave{\theta^2 v}$ and $-\ave{v^2\theta}$. The values are scaled by $u_\tau^3$, $u_\tau T_\tau^2$, or $u_\tau^2 T_\tau$. }
     \label{fig:uuv}
\end{figure}

We focus on the spatial turbulent transport of the Reynolds stresses and temperature-related statistics.
Figure~\ref{fig:uuv} presents the profiles of the triple velocity correlations $\ave{u^2v}$ and $-\ave{uv^2}$ and those of the triple velocity-temperature correlations $\ave{\theta^2 v}$ and $-\ave{v^2\theta}$. As for the physical interpretations of these third-order statistics, $\ave{u^2 v}$ and $\ave{\theta^2 v}$ represent respectively the wall-normal spatial fluxes of the velocity fluctuation $\ave{u^2}$ and the temperature fluctuation $\ave{\theta^2}$, which are caused by turbulent fluid motions. Similarly, $-\ave{u v^2}$ and $-\ave{v^2 \theta}$ indicate the turbulent spatial fluxes of the Reynolds shear stress $-\ave{uv}$ and the turbulent heat transfer $-\ave{v\theta}$, respectively. As discussed in Section~\ref{sec:analysis}, these third-order statistics represent the spatial transport effect caused by interactions between different scales. As shown in Fig.~\ref{fig:uuv}, the profiles of the triple correlations are all qualitatively similar, indicating transport further towards the wall in the near-wall region and transport towards the channel centre in the far-wall region. 

\begin{figure}[t!]
    \centering
    \includegraphics[width=1\hsize]{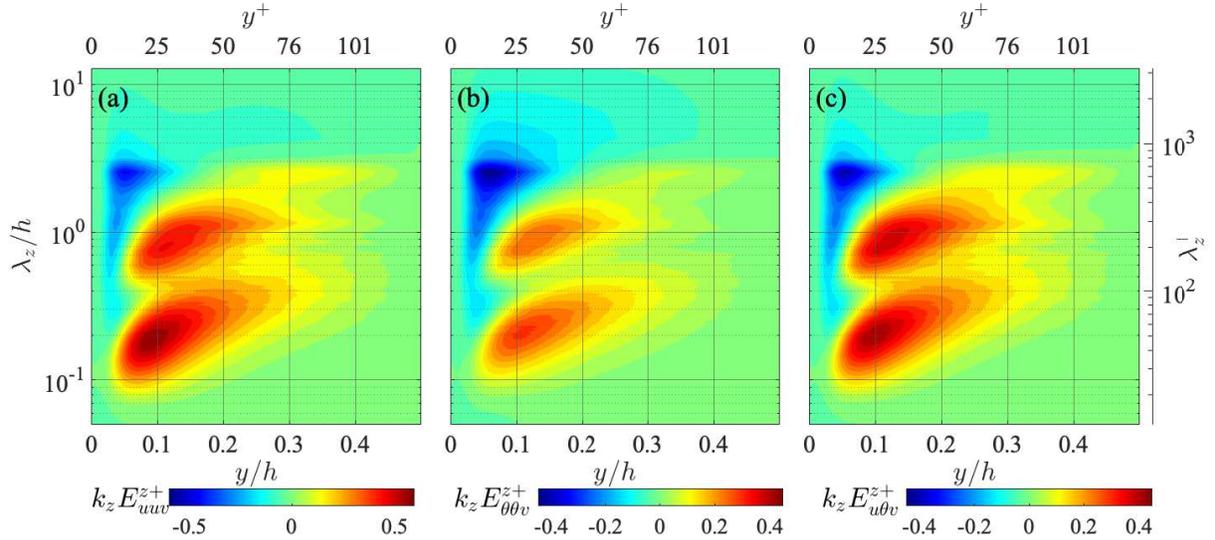}
    \caption{Spanwise one-dimensional premultiplied spectra of turbulent spatial fluxes of (a) streamwise turbulent energy~$k_z E^z_{uuv}$, (b) temperature fluctuation~$k_z E^z_{\theta\theta v}$, and (c) temperature-velocity correlation $k_z E^z_{u\theta v}$, presented in the same manner as in Fig.~\ref{fig:ez}. The values are scaled by $u_\tau^3$, $u_\tau T_\tau^2$, and $u_\tau^2 T_\tau$, respectively. }
    \label{fig:Ezuuv_vvt}
\end{figure}

Figure~\ref{fig:Ezuuv_vvt} presents the distributions of the spanwise one-dimensional spectra of the turbulent energy transports $E^z_{uuv}$, $E^z_{\theta \theta v}$, and $E^z_{u\theta v}$ in the premultiplied form. These spectra represent the spectral contents of the triple correlations $\ave{u^2v}$, $\ave{\theta^2 v}$, and $\ave{u \theta v}$ presented in Fig.~\ref{fig:uuv}. As shown in Fig.~\ref{fig:Ezuuv_vvt}, these spectra of spatial energy fluxes show similar distributions, indicating that the negative (i.e., towards the wall) spatial transport in the near-wall region mainly occurs at the largest wavelength $\lambda_z/h \approx 2.5$, while the positive (towards the channel centre) transport occurs at two smaller scales: the middle scale roughly at $\lambda_z/h \approx 0.8$ and relatively small wavelengths around $\lambda_z/h \approx 0.2$ ($\lambda_z^+ \approx 50$). As the wavelength of the negative $k_z E^z_{uuv}$ peak at the largest scale $\lambda_z/h = 2.5$ coincides with the typical spanwise spacing of $u$-streaks of the very-large-scale structure, the large-scale negative peak presumably represents the turbulence transports in which the fluctuating velocities and temperature are carried towards the wall vicinity by the secondary fluid motions of the very-large-scale structures. As for the positive $k_z E^z_{uuv}$ peak at a relatively small scale $\lambda_z^+ \approx 50$, this wavelength fairly matches the peak location of the premultiplied wall-normal turbulent energy spectrum $k_z E^z_{vv}$ in the near-wall region~\cite{tsukahara_2006}, which is likely related to the coherent fluid motions in the near-wall region. Therefore, it can be inferred that this positive $k_z E^z_{uuv}$ peak indicates the turbulent energy transport driven by the wall-normal fluid motions by the near-wall coherent structures. On the other hand, the wavelength of the positive $k_z E^z_{uuv}$ peak on the larger-scale side, $\lambda_z/h \approx 0.8$ ($\lambda_z^+ \approx 100$), is on the order of the full channel height $h$, but any corresponding characteristic length scale of coherent fluid motion is not found. As this scale lies in the middle wavelength range between the length scales of the very-large-scale and near-wall structures, this positive $k_z E^z_{uuv}$ peak may represent the energy transport caused by interactions between these inner and outer coherent structures. 

\begin{figure}
 \centering
     \includegraphics[width=0.7\hsize]{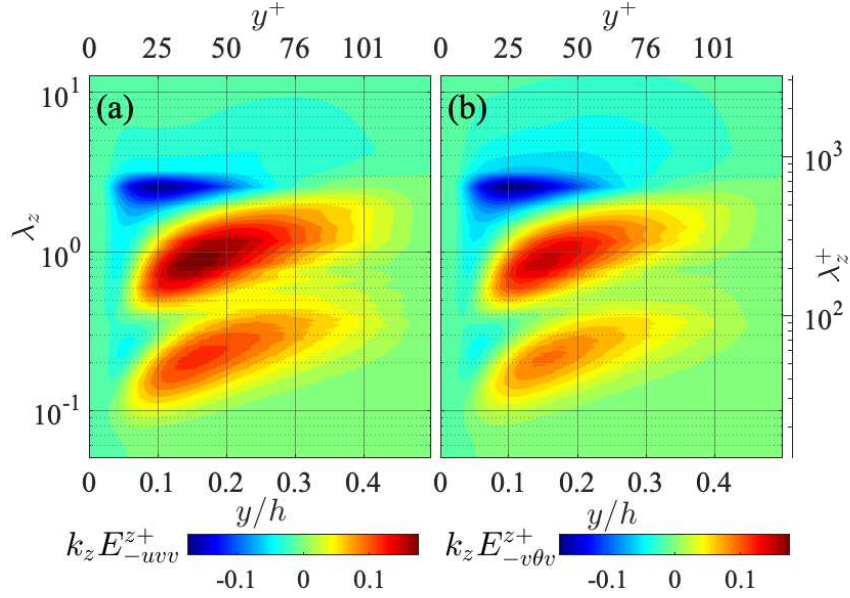}
     \caption{Spanwise one-dimensional premultiplied spectra of turbulent spatial fluxes of (a) Reynolds shear stress $E^z_{-uvv}$ and (b) the turbulent heat flux $E^z_{-v \theta v}$, presented in the same manner as in Fig.~\ref{fig:ez}. The values are scaled by $u_\tau^3$ and $u_\tau^2 T_\tau$, respectively. }
     \label{fig:Ezuvv_vtv}
\end{figure}

Figure~\ref{fig:Ezuvv_vtv} presents the distributions of the spatial flux spectra of the Reynolds shear stress $E^z_{-uvv}$ and  the turbulent heat transfer $E^z_{-v\theta v}$. Similarly to other spatial flux spectra presented in Fig.~\ref{fig:Ezuuv_vvt}, $E^z_{-uvv}$ and $E^z_{-v\theta v}$ are the spectral contents of the triple correlations $-\ave{uv^2}$ and $-\ave{v^2\theta}$ presented in Fig.~\ref{fig:uuv}, indicating the wall-normal fluxes of the Reynolds shear stress $-\ave{uv}$ and the turbulent heat flux $-\ave{v\theta}$ at each wavelength $\lambda_z$, respectively. As shown in the figure, these spatial flux spectra of the cross correlations present qualitatively similar distributions to those of the spatial turbulent energy fluxes, such as $E^z_{uuv}$ and $E^z_{\theta \theta v}$ presented in Fig.~\ref{fig:Ezuuv_vvt}, indicating the peak of negative transport at the largest wavelength $\lambda_z/h\approx 2.5$ near the wall and the positive transport peaks at two smaller wavelengths in the relatively far-wall region. This tendency is in contrast to the interscale fluxes in that the interscale fluxes of the cross correlations $Tr^z_{-uv}$ and $Tr^z_{-v\theta}$ show clearly different behaviours from those of turbulent energies such as $Tr^z_{uu}$ and $Tr^z_{\theta \theta}$, as observed in Figs.~\ref{fig:trzuu_tt} and \ref{fig:trzuv_vt}. 

As described so far, the spectral distributions of the fluctuating velocity and temperature fields have been investigated with respect to their interscale and spatial fluxes, and it has been observed that the temperature-related statistics essentially show similar distributions to the corresponding turbulence statistics, indicating a close similarity between the fluctuating velocity and temperature fields. In the next section, the spectral budgets of the transport equations of the temperature-related statistics are investigated further in detail. 

\subsection{Budget Analysis of Spectral Transport Equations}

\begin{figure}
    \hspace{-0.9cm}
    \includegraphics[width=1\hsize]{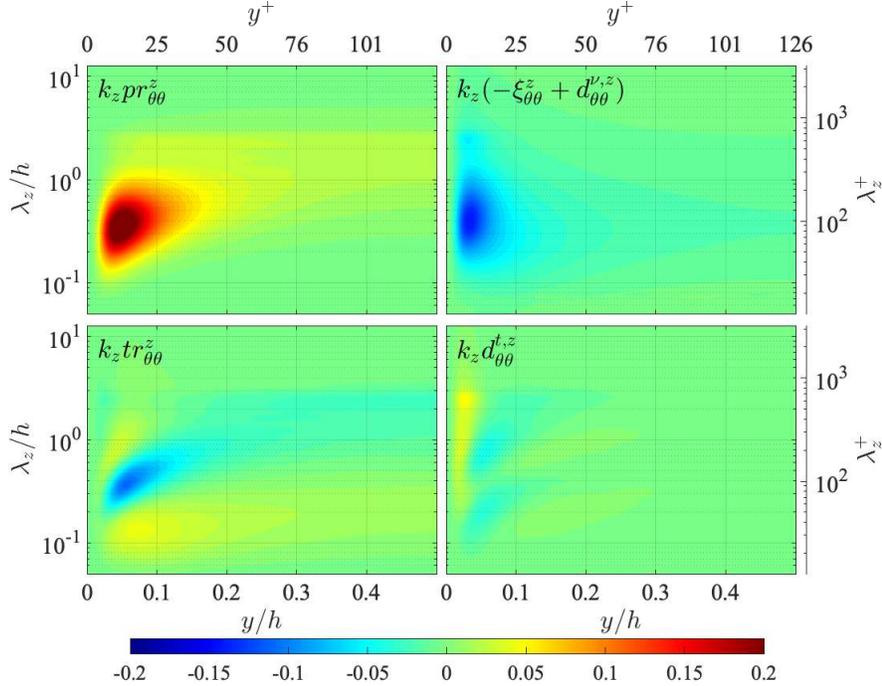}
    \caption{Space-wavelength ($y$-$\lambda_z$) diagrams of terms in the transport equation of the temperature fluctuation spectrum $E^z_{\theta \theta}$: (top left) production $pr^z_{\theta \theta}$; (top right) viscous terms $\xi^z_{\theta \theta}+d^{\nu,z}_{\theta \theta}$; (bottom left) interscale transport $tr^z_{\theta \theta}$; (bottom right) turbulent diffusion $d^{t,z}_{\theta \theta}$. The terms are presented in the premultiplied form, and the values are scaled by $u_\tau^2 T_\tau^2 /\nu$. }
    \label{fig:ttbdgt}
\end{figure}

Now, we investigate the transport budget of the temperature-related spectra $E^z_{\theta \theta}$, $E^z_{u\theta}$, and $E^z_{-v\theta}$ in detail. The transport budgets of the Reynolds stress spectra are also given in Apendix~\ref{sec:rsbdgt}, and compared to these temperature-related spectra transport when necessary for discussion. Figure~\ref{fig:ttbdgt} presents the distributions of terms in the transport equation of the temperature fluctuation spectrum $E^z_{\theta \theta}$:
\begin{align}
\left( \pd{}{t}+U_k\pd{}{x_k} \right) E^z_{\theta \theta} = pr^z_{\theta \theta} - \xi^z_{\theta \theta} + d^{\nu,z}_{\theta \theta} + d^{t,z}_{\theta \theta} + tr^z_{\theta \theta}.
\end{align}
As shown in the top-left panel, the temperature fluctuation is produced by $pr^z_{\theta \theta}$ mainly at a relatively small wavelength of $\lambda_z^+ \approx 100$ in the near-wall region, which roughly corresponds to the $y$-position and the spanwise length scale of the coherent structure in the near-wall region. This tendency is quite similar to the distribution of  the streamwise turbulent energy production spectrum $pr^z_{uu}$ given in Fig.~\ref{fig:uubdgt}, see Appendix~\ref{sec:rsbdgt}. The similarity between these productions can be understood by noting that $pr^z_{\theta \theta}$ and $pr^z_{uu}$ depend on the cospectra $E^z_{-v\theta}$ and $E^z_{-uv}$, respectively, as 
\begin{align}
pr^z_{\theta \theta} = E^z_{-v\theta} \frac{\mathrm{d} \varTheta}{\mathrm{d} y}, \quad
pr^z_{uu} = E^z_{-uv} \frac{\mathrm{d} U}{\mathrm{d} y},
\end{align}
and the distributions of $E^z_{-v\theta}$ and $E^z_{-uv}$ and the profiles of $U$ and $\varTheta$ are similar to each other, respectively, as presented in Figs.~\ref{fig:meanUT} and \ref{fig:uv_vtz}. 

The distribution of the interscale transport $tr^z_{\theta \theta}$ presents the consequential energy gain/loss by the interscale energy flux $Tr^z_{\theta \theta}$ shown in Fig.~\ref{fig:trzuu_tt} as $tr^z_{\theta \theta}=-\partial Tr^z_{\theta \theta}/\partial k_z$, and it is shown in Fig.~\ref{fig:ttbdgt} that the energy is removed from the region around the peak of $pr^z_{\theta \theta}$ and transported towards both larger and smaller $\lambda_z$ ranges in the near-wall region by the interscale energy transport $tr^z_{\theta \theta}$, while in the channel-core region the energy is mainly transported in the forward direction from around $\lambda_z/h \approx 2.5$ to a smaller $\lambda_z$ range.  

The distribution of the spectral turbulent diffusion $d^{t,z}_{\theta \theta}$ represents the energy gain/loss by the turbulent spatial flux $E^z_{\theta \theta v}$ presented in Fig.~\ref{fig:Ezuuv_vvt}, as $d^{t,z}_{\theta \theta} = -\partial E^z_{\theta \theta v}/\partial y$. Its contribution is relatively small as compared to the other terms, but it can be seen that the energy is supplied to the wall vicinity at a large wavelength $\lambda_z/h \approx 2.5$, and the energy transport towards the channel central region is also observed at two relatively small wavelength ranges, corresponding to the distribution of the turbulent spatial flux $E^z_{\theta \theta v}$. The energy produced by $pr^z_{\theta \theta}$ and transported by $tr^z_{\theta \theta}$ and $d^{t,z}_{\theta \theta}$ is eventually dissipated by the viscous dissipation $\xi^z_{\theta \theta}$. Such distributions of the terms in the $E^z_{\theta \theta}$ transport equation described above are quite similar to those of the corresponding terms in the $E^z_{uu}$ transport equation given in Fig.~\ref{fig:uubdgt}. 

The clear difference between the transport equations of the temperature fluctuation spectrum $E^z_{\theta \theta}$ and the streamwise turbulent energy $E^z_{uu}$ is that no pressure-related term exists in the $E^z_{\theta \theta}$ transport equation, whereas in the $E^z_{uu}$ transport the pressure-strain cospectrum $\pi^z_{uu}$ plays an important role to redistribute energy from $E^z_{uu}$ to other components of turbulent energy spectra. Despite such a distinct difference, the distributions of the terms in the $E^z_{\theta \theta}$ equation are quite similar to those of the corresponding terms of the $E^z_{uu}$ transport equation (compare Figs.~\ref{fig:ttbdgt} and \ref{fig:uubdgt}). In the following, the budget balances of the $E^z_{\theta \theta}$ and $E^z_{uu}$ equations are compared in more detail. 

\begin{figure}[t!]
    \centering
    \includegraphics[width=0.7\hsize]{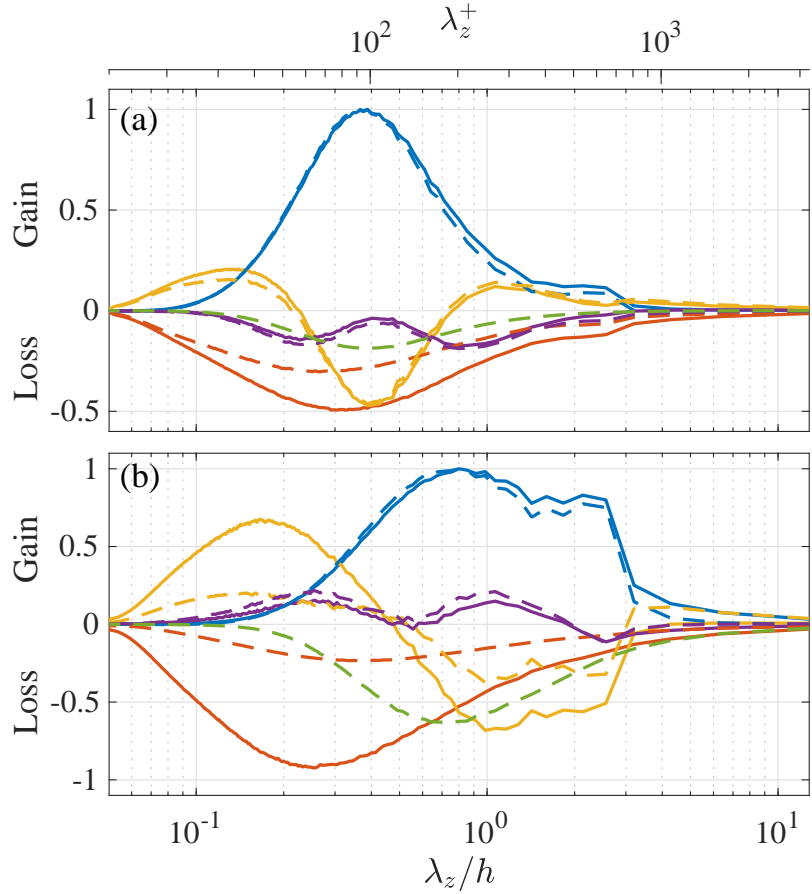}
    \caption{Detailed budget balance of the $E^z_{\theta \theta}$ and $E^z_{uu}$ transport equation at (a) a near-wall location $y^+=16$ and (b) the quarter height of the channel $y/h=0.25$: (blue) production; (red) viscous dissipation$+$viscous diffusion; (yellow) interscale transport; (purple) turbulent diffusion; (light green) pressure-strain cospectrum~$\pi^z_{uu}$. The solid and dashed lines represent terms in the $E^z_{\theta \theta}$ and $E^z_{uu}$ equations, respectively. The terms are given in the premultiplied form, and the values are scaled by the peak value of each production term. } 
    \label{fig:ttuubdgt}
\end{figure}

Figure~\ref{fig:ttuubdgt} presents detailed balance between the terms of the $E^z_{\theta \theta}$ and $E^z_{uu}$ equations at a near-wall location $y^+=16$ and the quarter height of the channel $y/h=0.25$ ($y^+ \approx 60$). The values are scaled by the maximum value of each production term. As shown in the panel~(a), the profiles of the production, interscale transport, and turbulent diffusion of the $E^z_{\theta \theta}$ and $E^z_{uu}$ transport equations show a good collapse in the near-wall region. The production terms indicate a significant energy gain at $\lambda_z^+ \approx 100$, as already pointed out above, and the interscale transport terms are shown to transfer the energy from this energy-producing wavelength to both larger and smaller $\lambda_z$ ranges, where the turbulent diffusion terms transfer some energy to other $y$-positions. As previously mentioned, there is no term in the $E^z_{\theta \theta}$ transport budget corresponding to the pressure-strain term $\pi^z_{uu}$ in the $E^z_{uu}$ transport, and the consequential difference between the budget balances of the $E^z_{\theta \theta}$ and $E^z_{uu}$ transport equation is mainly manifested by the viscous terms. As there is no  energy absorption by the pressure-strain cospectrum in the $E^z_{\theta \theta}$ budget, the viscous terms $\xi^z_{\theta \theta}+d^{\nu,z}_{\theta \theta}$ show a clearly larger contribution than $\xi^z_{uu}+d^{\nu,z}_{uu}$ to compensate for the absence of the pressure term and, thus, the transport equation is balanced. 

At the quarter height of the channel $y/h=0.25$, as shown in Fig.~\ref{fig:ttuubdgt}(b), the profiles of the production terms collapse similarly to those in the near-wall region, but the $\lambda_z$  range where energy is dissipated by viscosity is significantly smaller than the energy-producing $\lambda_z$ range. Hence, the interscale transport terms bridge these energy-producing and -dissipating $\lambda_z$ ranges by forward energy transfer and do not exhibit reversed energy transfer, unlike in the near-wall region. It is also observed that the production terms present not only the largest peak of energy gain at $\lambda_z^+\approx 200$ but also a secondary peak at $\lambda_z/h \approx 2.5$, which corresponds to the very-large-scale structures. The turbulent diffusion terms present energy gains at two wavelength ranges, which represents energy supply from the near-wall region (see the $d^{t,z}_{\theta \theta}$ and $d^{t,z}_{uu}$ distributions in Figs.~\ref{fig:ttbdgt} and \ref{fig:uubdgt}). The contribution of the pressure-strain correlation $\pi^z_{uu}$ in the $E^z_{uu}$ transport equation is relatively large compared to that in the near-wall location shown in the panel~(a), and the interscale transport $tr^z_{\theta \theta}$ as well as the viscous terms $\xi^z_{\theta \theta}+d^{\nu,z}_{\theta \theta}$ show significantly larger contributions than the counterparts in the $E^z_{uu}$ equation to compensate for the absence of the energy absorption by $\pi^z_{uu}$. Thus, both in the near- and far-wall regions of the channel, the energy productions in the $E^z_{\theta \theta}$ and $E^z_{uu}$ transports are quite similar to each other, but more energy is dissipated by the viscous terms in the $E^z_{\theta \theta}$ transport, compensating for the absence of the energy loss by the pressure-related term.

\begin{figure}
    \centering
    \includegraphics[width=1\hsize]{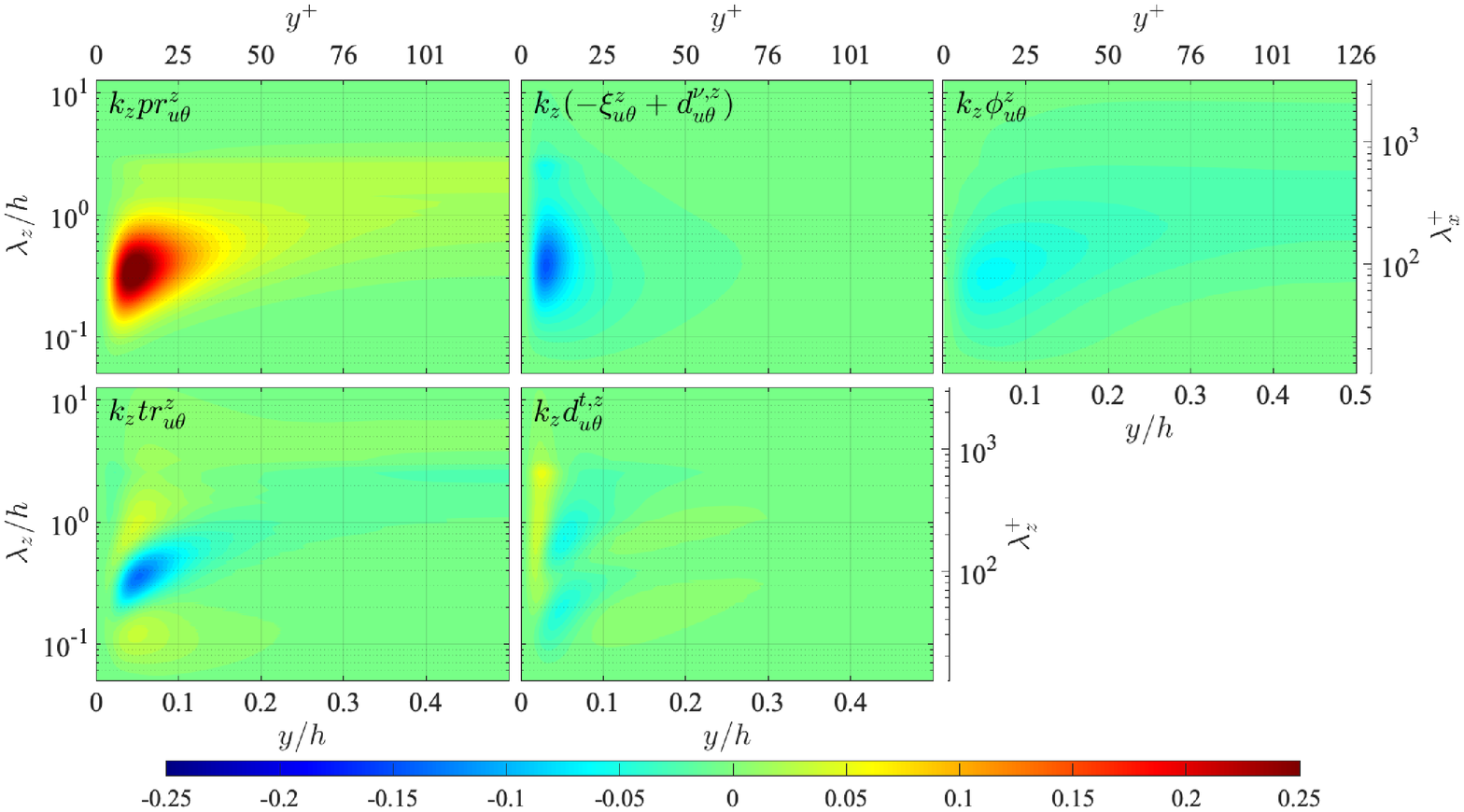}
    \caption{Distribution of terms in the transport equation of the velocity-temperature cospectrum $E^z_{u\theta}$ presented in the same manner as in Fig.~\ref{fig:ttbdgt}: (top left) production $k_z pr^z_{u\theta}$; (top centre) viscous terms $k_z(\xi^z_{u\theta}+d^{\nu,z}_{u\theta})$; (top right)~pressure term $k_z \phi^z_{u\theta}$; (bottom right)~interscale transport $tr^z_{u\theta}$; (bottom centre)~turbulent diffusion $d^{t,z}_{u\theta}$. The values are scaled by $u_\tau^3 T_\tau /\nu$.}
    \label{fig:utbdgt}
\end{figure}

The transport equation of the velocity-temperature cospectrum $E^z_{u\theta}$ is written as 
\begin{align}
\left( \pd{}{t}+U_k\pd{}{x_k} \right) E^z_{u\theta} = pr^z_{u\theta} - \xi^z_{u\theta} + d^{\nu,z}_{u\theta} + \phi^z_{u\theta} + d^{t,z}_{u\theta} +tr^z_{u\theta},
\end{align}
and the distributions of the terms in the right-hand side are presented in Fig.~\ref{fig:utbdgt}. It is clearly seen here that the $E^z_{u\theta}$ transport equation has a pressure term $\phi^z_{u\theta}$ and it essentially functions as an energy sink throughout the channel, similarly to the pressure-strain energy redistribution term $\pi^z_{uu}$ in the $E^z_{uu}$ transport equation. It is also noteworthy that the production of the velocity-temperature cospectrum $pr^z_{u\theta}$ depends on both the Reynolds shear stress spectra $E^z_{-uv}$ and the turbulent heat flux spectra $E^z_{-v\theta}$ as
\begin{align}
pr^z_{u\theta} = E^z_{-v\theta} \frac{\mathrm{d} U}{\mathrm{d}y} + E^z_{-uv} \frac{\mathrm{d} \varTheta}{\mathrm{d} y},
\end{align}  
unlike the productions of other spectra, such as $pr^z_{uu}$ and $pr^z_{\theta \theta}$, which only depend on either $E^z_{-uv}$ or $E^z_{-v\theta}$. The distribution of $pr^z_{u\theta}$ is, however, similar to those of $pr^z_{uu}$ and $pr^z_{\theta \theta}$, as the distributions of $E^z_{-uv}$ and $E^z_{-v\theta}$ are similar to each other and so are the profiles of $U$ and $\varTheta$. Each term of the $E^z_{u\theta}$ transport equation presented in Fig.~\ref{fig:utbdgt} shows a similar distribution to the corresponding term in the $E^z_{uu}$ transport equation shown in Fig.~\ref{fig:uubdgt}. Since the $E^z_{u\theta}$ transport equation has a pressure-related term, unlike the $E^z_{\theta \theta}$ equation, the budget balance is even closer to the transport of the streamwise turbulent energy $E^z_{uu}$ than that of the $E^z_{\theta \theta}$ transport. 

\begin{figure}
    \centering
    \includegraphics[width=1\hsize]{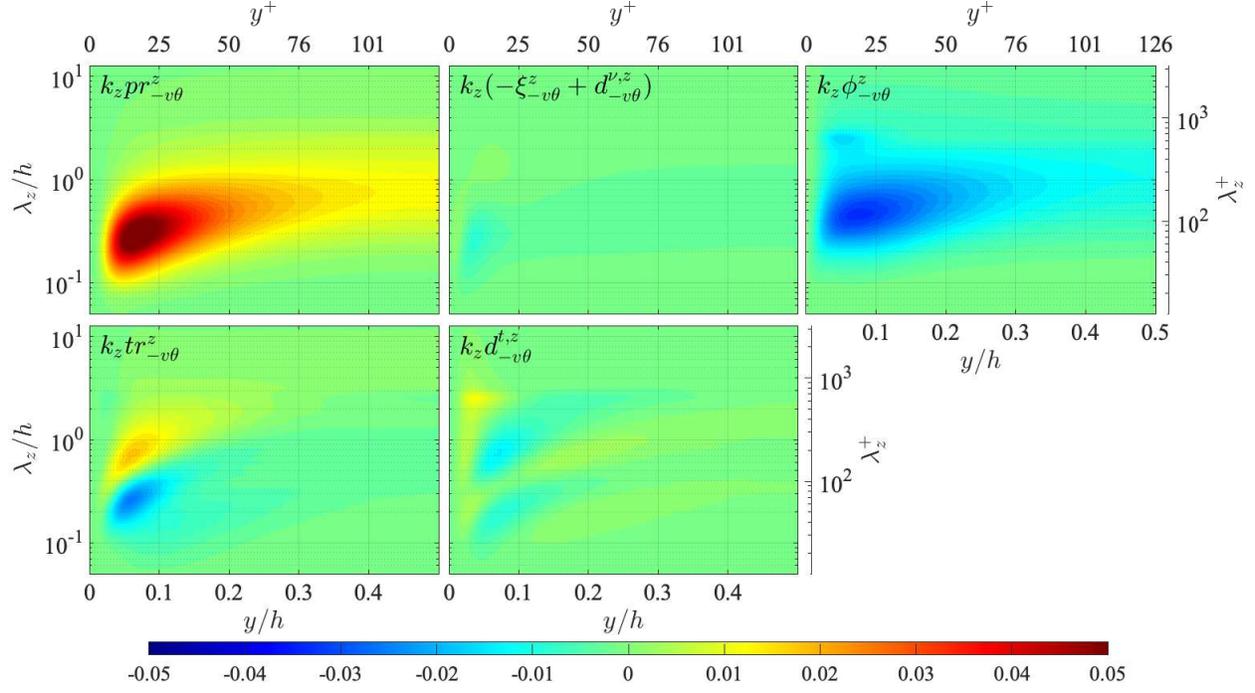}
    \caption{Distribution of terms in the transport equation of the turbulent heat flux spectrum $E^z_{-v\theta}$ presented in the same manner as in Fig.~\ref{fig:utbdgt}: (top left)~production $k_z pr^z_{-v\theta}$; (top centre)~viscous terms $k_z(\xi^z_{-v\theta}+d^{\nu,z}_{-v\theta})$; (top right)~pressure term $k_z \phi^z_{-v\theta}$; (bottom left)~interscale transport $tr^z_{-v\theta}$; (bottom centre)~turbulent diffusion $d^{t,z}_{-v\theta}$.  The values are scaled by $u_\tau^3 T_\tau/\nu$.}
    \label{fig:vtbdgt}
\end{figure}

Now, we shed light on the transport of the turbulent heat flux spectrum $E^z_{-v\theta}$. The transport equation of $E^z_{-v\theta}$ is written as
\begin{align}
\left( \pd{}{t}+U_k\pd{}{x_k} \right) E^z_{-v\theta} = pr^z_{-v\theta} - \xi^z_{-v\theta} + d^{\nu,z}_{-v\theta} + \phi^z_{-v\theta} + d^{t,z}_{-v\theta} +tr^z_{-v\theta},
\end{align}
and the distributions of the terms on the right-hand side are presented in Fig.~\ref{fig:vtbdgt}. As shown here, $E^z_{-v\theta}$ is produced by the production term $pr^z_{-v\theta}$ mainly in the near-wall region at wavelengths around $\lambda_z^+ \approx 100$, similarly to the other energy spectra $pr^z_{\theta \theta}$ and $pr^z_{u\theta}$. The energy gain by the production $pr^z_{-v\theta}$ is mainly balanced by the pressure term $\phi^z_{-v\theta}$, and the viscous terms do not play an important role. 

The turbulent transport terms $tr^z_{-v\theta}$ and $d^{t,z}_{-v\theta}$ are shown to partly transport $E^z_{-v\theta}$ in scale and space, respectively. The interscale transport $tr^z_{-v\theta}$ presents the energy gain/loss by the inverse interscale flux $Tr^z_{-v\theta}$ investigated in Fig.~\ref{fig:trzuv_vt}, and the distribution shows that the turbulent heat flux $-\ave{v\theta}$ is mainly transported from wavelengths around $\lambda_z/h \approx 0.25$ ($\lambda_z^+ \approx 65$) to a larger $\lambda_z$ range around $\lambda_z/h \approx 0.75$ ($\lambda_z^+ \approx 190$) in the near-wall region. The distribution of $d^{t,z}_{-v\theta}$ indicates weak spatial transports similarly to $d^{t,z}_{\theta \theta}$ and $d^{t,z}_{u \theta}$, where $-\ave{v\theta}$ is supplied to the wall vicinity from the channel-core region at the largest wavelengths $\lambda_z/h\approx2.5$ and at a smaller $\lambda_z$ range $-\ave{v\theta}$ is transported from the near- to far-wall region at two different wavelengths. 

The tendencies of the terms in the $E^z_{-v\theta}$ transport equation described above are quite similar to those of the corresponding terms in the transport equation of the Reynolds shear stress spectrum $E^z_{-uv}$, as can be seen by comparing Fig.~\ref{fig:vtbdgt} with Fig.~\ref{fig:uvbdgt} in Appendix~\ref{sec:rsbdgt}. It is particularly noteworthy here that the productions of $E^z_{-v\theta}$ and $E^z_{-uv}$ are both dependent on the wall-normal turbulent energy spectrum $E^z_{vv}$ as
\begin{align}
pr^z_{-v\theta} = E^z_{vv} \frac{\mathrm{d} \varTheta}{\mathrm{d}y}, \quad 
pr^z_{-uv} = E^z_{vv} \frac{\mathrm{d} U}{\mathrm{d}y}, \label{eq:prvt_uv}
\end{align}
which indicates that the Reynolds shear stress $-\ave{uv}$ and the turbulent heat flux $-\ave{v\theta}$ are produced at the same scale.  

\begin{figure}[t!]
    \centering
    \includegraphics[width=0.7\hsize]{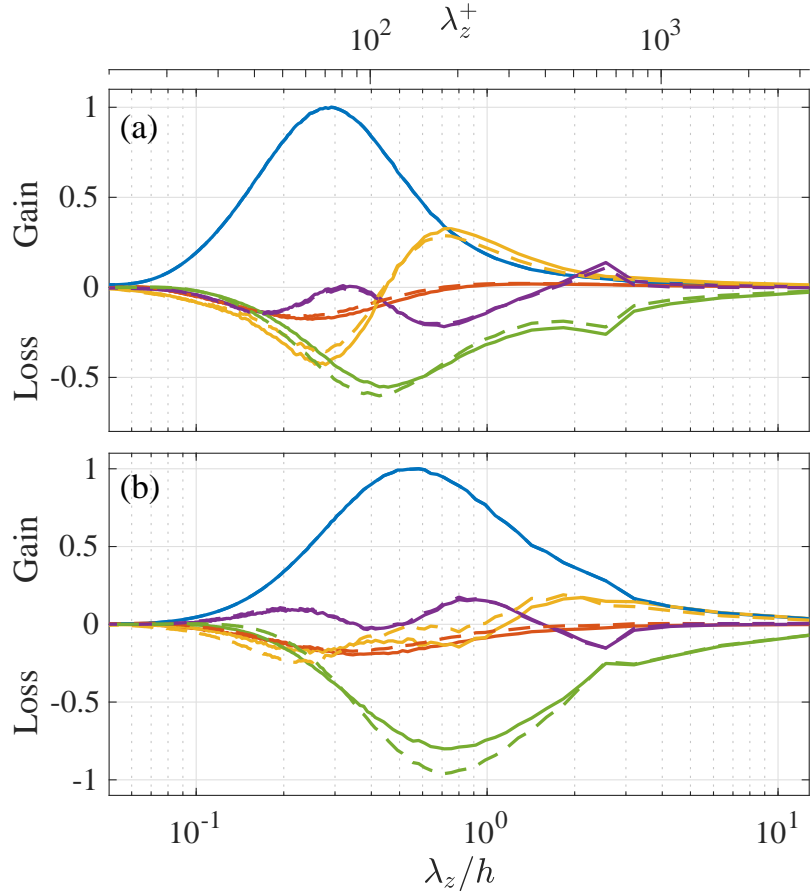}
    \caption{Detailed budget balance of the $E^z_{-v\theta}$ and $E^z_{-uv}$ transport equation at (a) a near-wall location $y^+=16$ and (b) the quarter height of the channel $y/h=0.25$ presented in the same manner as in Fig.~\ref{fig:ttuubdgt}: (blue) production; (red) viscous dissipation$+$viscous diffusion; (yellow) interscale transport; (purple) turbulent diffusion; (light green) pressure term.} 
    \label{fig:vtuvbdgt}
\end{figure}

Figure~\ref{fig:vtuvbdgt} presents a detailed comparison of the transport budgets of $E^z_{-v\theta}$ and $E^z_{-uv}$ at a near-wall location $y^+=16$ and the quarter height of the channel in the same manner as in Fig.~\ref{fig:ttuubdgt}. Note here that in both panels the profiles of the productions $pr^z_{-v\theta}$ and $pr^z_{-uv}$ are exactly on top of each other since their profiles are similar, as shown by Eq.~(\ref{eq:prvt_uv}). As shown here, at both wall-normal locations, the spectral budget balances of the $E^z_{-v\theta}$ and $E^z_{-uv}$ transport are quite similar to each other. It is also shown at both wall-normal locations that the peaks of the pressure terms are located at a larger $\lambda_z$ than those of the production terms in both $E^z_{-v\theta}$ and $E^z_{-uv}$ transport, which indicates that the turbulent heat flux $-\ave{v\theta}$ and the Reynolds shear stress $-\ave{uv}$ are dissipated by the effect of pressure at larger spanwise wavelengths than those at which they are produced by the mean temperature or velocity gradients. Such tendencies are in contrast to the transport of the energy spectra such as $E^z_{uu}$ and $E^z_{\theta \theta}$, where the energy is dissipated basically at smaller scales than produced. This is attributable to the contribution by the interscale transport terms $tr^z_{-v\theta}$ and $tr^z_{-uv}$ to transport the energy from smaller to larger scales.  

It is interesting to note that the agreements between the turbulent diffusion terms $d^{t,z}_{-v\theta}$ and $d^{t,z}_{-uv}$ is particularly good while a slight difference can be found between the interscale transport terms $tr^z_{-v\theta}$ and $tr^z_{-uv}$, although both the turbulent diffusion and interscale transport represent the transport effect caused by nonlinear interactions between different scales. Such good agreement between the turbulent diffusion terms are also observed between the $E^z_{\theta \theta}$ and $E^z_{uu}$ transports, as already pointed out with Fig.~\ref{fig:ttuubdgt}. 

\section{Discussion and Concluding Remarks}

In the previous section, the spectral budgets of the temperature-related spectra, such as $E^z_{\theta \theta}$, $E^z_{u \theta}$, and $E^z_{-v \theta}$, were compared with those of the corresponding Reynolds stress spectra, $E^z_{uu}$ and $E^z_{-uv}$, and close analogy between those turbulence transports was indicated. Of particular interest is the close similarity between the spectral transports of the Reynolds shear stress $-\ave{uv}$ and the turbulent heat flux $-\ave{v\theta}$, where inverse interscale transport from a smaller to larger $\lambda_z$ is observed. 

The turbulent diffusion and interscale transport are two different aspects of nonlinear multi-scale interactions of turbulence, and the physical phenomena these spatial and interscale transport terms represent is of great interest. As shown in Fig.~\ref{fig:vtuvbdgt}(a), the turbulent diffusions $d^{t,z}_{-v\theta}$ and $d^{t,z}_{-uv}$ are shown to remove energy from the near-wall location in two $\lambda_z$ ranges around $\lambda_z^+ \approx 50$ and $180$, and the removed $-\ave{v\theta}$ and $-\ave{uv}$ are partly transported towards the central region of the channel, as indicated by the profiles of $d^{t,z}_{-v\theta}$ and $d^{t,z}_{-uv}$ at $y/h=0.25$ given in Fig.~\ref{fig:vtuvbdgt}(b). In particular, the energy gains by $d^{t,z}_{-v\theta}$ and $d^{t,z}_{-uv}$ at $y/h=0.25$ are shown to balance well with the energy loss by the interscale energy transport $tr^z_{-v\theta}$ and $tr^z_{-uv}$, which suggests that  the energy ($-\ave{v\theta}$ and $-\ave{uv}$) spatially transferred from the near-wall region to this wall-normal location at relatively small $\lambda_z$ is further transferred to a larger $\lambda_z$ range. Kawata \& Alfredsson~\cite{kawata_2018} experimentally observed similar Reynolds shear stress transport from smaller $\lambda_z$ near the wall to larger $\lambda_z$ in the channel-core region in turbulent plane Couette flow and conjectured that it may represent the influences of the near-wall smaller-scale coherent structures on the very-large-scale structures, which is similar to the concept of a co-supporting cycle of the inner and outer structures proposed by Toh \& Itano~\cite{toh_2005}. 

It should be mentioned here that the inverse interscale transport of the Reynolds shear stress and turbulent heat flux described so far are observed based on one-dimensional spanwise-Fourier-mode analysis, and therefore, the interscale energy transfer in the streamwise length scales is not investigated. In fact, such reversed interscale transfers are not observed by the analysis based on the streamwise Fourier modes. In a  numerical simulation by Kawata \& Tsukahara~\cite{kawata_2021}, spectral budget analysis was performed based on both one-dimensional streamwise and spanwise Fourier modes, and it was shown that the interscale transfers of the turbulent energy and Reynolds shear stress in the streamwise wavenumber direction are basically forward transfers (i.e., from larger to smaller scales). Lee \& Moser~\cite{lee_2019} performed two-dimensional Fourier-mode analysis on the turbulent energy transport and showed that the energy transport between wavenumbers with the same magnitude but different directions, which may not represent energy transfer between really different length scales, can be observed as inverse interscale energy transfers through one-dimensional Fourier mode analysis. These observations suggest that the inverse energy transfers in the spanwise length scales may not simply be interpreted as interaction from smaller- to larger-scale structures. 

Some recent studies, on the other hand, investigated the relation between the inner-outer interaction in wall turbulence and the inverse interscale energy transfers in the spanwise length scales in detail. Doohan {\it et al}.~\cite{hwang_2021} investigated the interactions between self-sustaining processes (SSPs) of coherent structures near and away from the wall and argued that energy transfer from the smaller-scale SSP leads to the formation of the wall-reaching part of streaks of the larger-scale SSP. Chiarini {\it et al.}~\cite{gatti_2021} also investigated spatial and interscale energy fluxes by analysing the anisotropic generalised Kolmogorov equation taking into account both the streamwise and spanwise length scale and observed some energy paths from smaller scales near the wall to larger scales away from the wall. Chan {\it et al.}~\cite{chin_2021} also observed the inverse interscale transport of the Reynolds shear stress in a turbulent boundary layer, and based on a detailed quadrant analysis they interpreted it as the net energy transfer from the small-scale ejection (Q2) and sweep (Q4) events to their large-scale counterparts. Thus, what the inverse interscale energy transports in the spanwise lengths scales is still a subject of intense debate. Nevertheless, the transport of the turbulent heat flux spectrum $E^z_{-v\theta}$ is found to be quite similar to that of the Reynolds shear stress spectrum $E^z_{-uv}$, which indicates a close scale-by-scale similarity between the turbulent momentum and heat transfers.

In the present study, the transport budgets of temperature-related spectra, such as the temperature-variance and turbulent heat flux spectra, were investigated in turbulent plane Couette flow with a passive-scalar heat transfer at the Reynolds number $Re_\tau = 126$ and the Prandtl number $Pr=0.71$ based on DNS data. It was found that the transport budgets of the temperature-related spectra  present  quite similar tendencies to those of the corresponding Reynolds stress spectra, including the spectral transport of the Reynolds shear stress and turbulent heat flux, where inverse interscale transfers are observed throughout the channel. It is also noteworthy that the distributions of the spatial flux spectra are all similar, regardless the spatial fluxes of turbulent energies such as $\ave{u^2}$ and $\ave{\theta^2}$ or those of cross correlations such as $-\ave{uv}$ and $-\ave{v\theta}$.  This is in contrast to that the interscale fluxes of $-\ave{uv}$ and $-\ave{v\theta}$ exhibit notably different behaviours from those of turbulent energies such as $\ave{u^2}$ and $\ave{\theta^2}$.  Our next task is to elucidate what physical phenomena these spatial and interscale turbulence transport represent. The present investigation is limited to single values of the Reynolds and Prandtl number, and therefore, the Reynolds- and Prandtl-number dependency of these interscale and spatial fluxes should also be investigated in the future studies. 

\appendix
\section{Spectral Budgets of the Reynolds Stress Transports} \label{sec:rsbdgt}

\begin{figure}
    \centering
    \includegraphics[width=1\hsize]{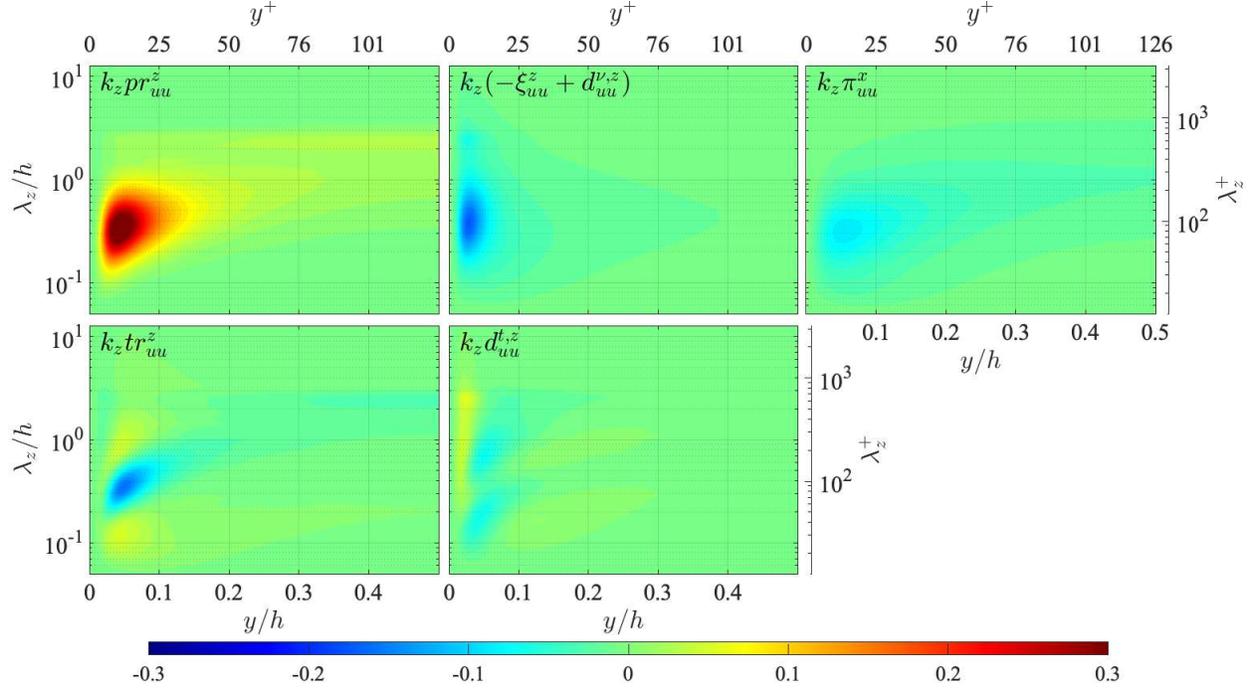}
    \caption{Distribution of terms in the transport equation of the streamwise turbulent energy spectrum $E^z_{uu}$, presented in the same manner as in Fig.~\ref{fig:utbdgt}: (top left)~production $k_z pr^z_{uu}$; (top centre)~viscous terms $k_z(\xi^z_{uu}+d^{\nu,z}_{uu})$; (top right)~pressure-strain energy redistribution $k_z \pi^z_{uu}$; (bottom left)~interscale transport $tr^z_{uu}$; (bottom centre)~turbulent diffusion $d^{t,z}_{uu}$. The values are scaled by $u_\tau^4/\nu$.}
    \label{fig:uubdgt}
\end{figure}

\begin{figure}[t]
    \centering
    \includegraphics[width=1\hsize]{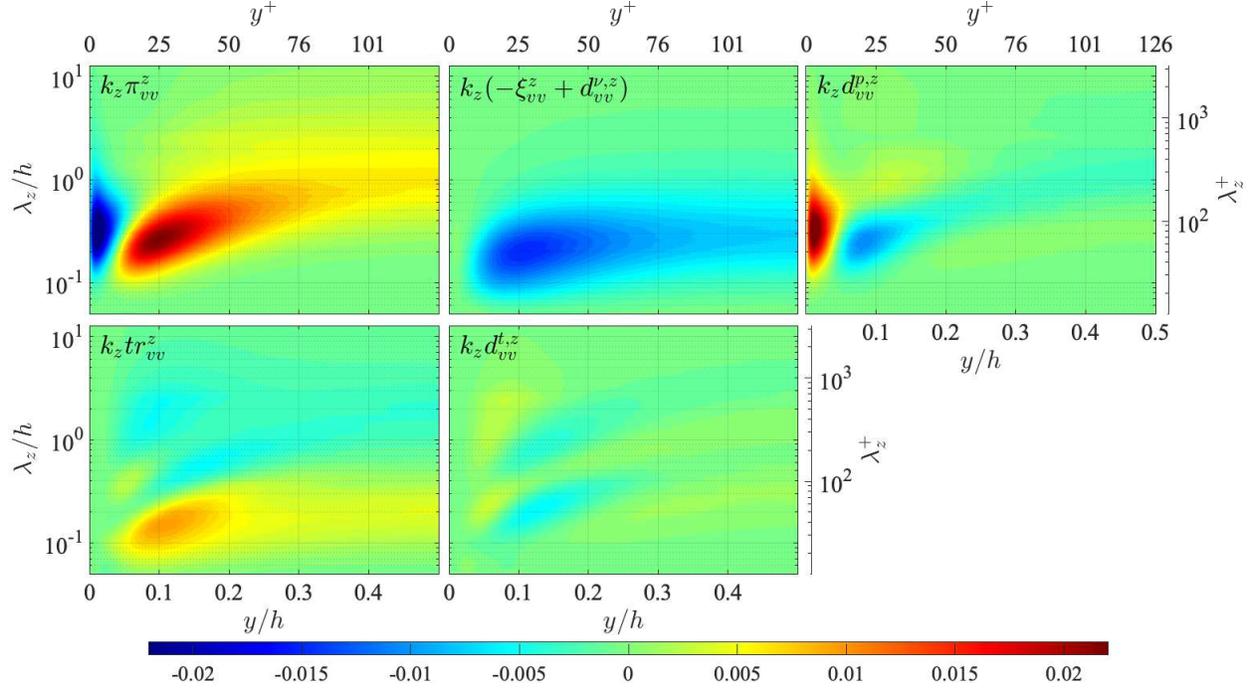}
    \caption{Distribution of terms in the transport equation of the wall-normal turbulent energy spectrum $E^z_{vv}$, presented in the same manner as in Fig.~\ref{fig:utbdgt}: (top left)~pressure-strain energy redistribution $k_z \pi^z_{ww}$; (top centre)~viscous terms $k_z(-\xi^z_{ww}+d^{\nu,z}_{ww})$; (top right) pressure diffusion $k_z d^{p,z}_{vv}$; (bottom left)~interscale transport $tr^z_{vv}$; (bottom right)~turbulent diffusion $d^{t,z}_{vv}$. The values are scaled by $u_\tau^4/\nu$.}
    \label{fig:vvbdgt}
\end{figure}

\begin{figure}[t]
\centering
    \hspace{-1.25cm}
    \includegraphics[width=1\hsize]{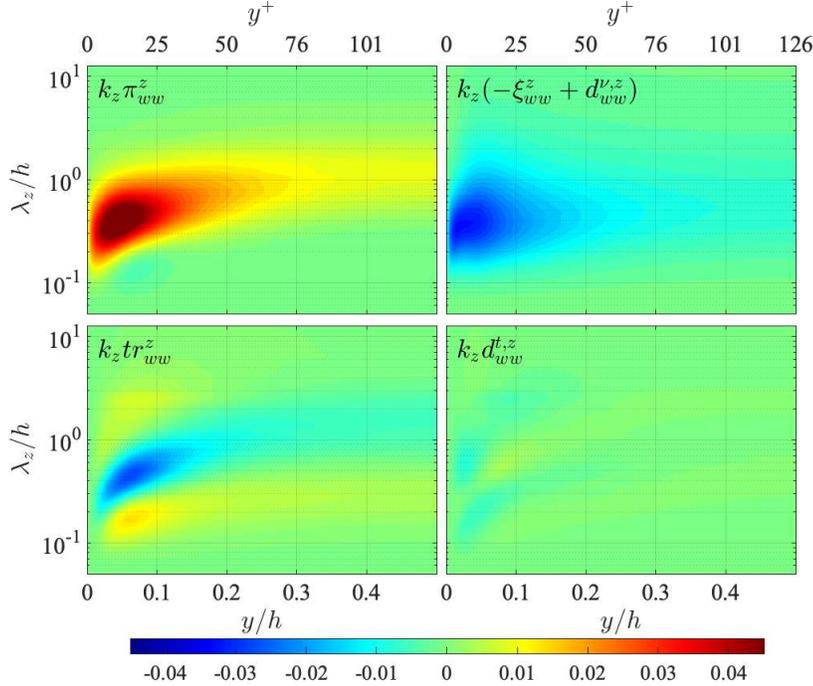}
    \caption{Distribution of terms in the transport equation of the spanwise turbulent energy spectrum $E^z_{ww}$, presented in the same manner as in Fig.~\ref{fig:utbdgt}: (top left)~pressure-strain energy redistribution $k_z \pi^z_{ww}$; (top right)~viscous terms $k_z(-\xi^z_{ww}+d^{\nu,z}_{ww})$;(bottom left)~interscale transport $tr^z_{ww}$; (bottom right)~turbulent diffusion $d^{t,z}_{ww}$. The values are scaled by $u_\tau^4/\nu$.}
    \label{fig:wwbdgt}
\end{figure}

Here, the transport budgets of the Reynolds stress spectra are presented for comparison to those of the temperature-related spectra given in Figs.~\ref{fig:ttbdgt}-\ref{fig:vtuvbdgt}. The transport equations of the streamwise turbulent energy $E^z_{uu}$, wall-normal turbulent energy $E^z_{vv}$, and spanwise turbulent energy $E^z_{ww}$ are, respectively, written as 
\begin{align}
\left( \pd{}{t}+U_k\pd{}{x_k} \right) E^z_{uu} &=  pr^z_{uu} - \xi^z_{uu} + d^{\nu,z}_{uu} + \pi^z_{uu} 
\hspace{0.9cm}+ d^{t,z}_{uu} + tr^z_{uu}, \\
\left( \pd{}{t}+U_k\pd{}{x_k} \right) E^z_{vv} &= \hspace{0.7cm} - \xi^z_{vv} + d^{\nu,z}_{vv} + \pi^z_{vv} + d^{p,z}_{vv}  + d^{t,z}_{vv} + tr^z_{vv}, \\
\left( \pd{}{t}+U_k\pd{}{x_k} \right) E^z_{ww} &= \hspace{0.7cm} - \xi^z_{ww} + d^{\nu,z}_{ww} + \pi^z_{ww} 
\hspace{0.7cm} + d^{t,z}_{ww} + tr^z_{ww},
\end{align}
and the distributions of each term on the right-hand side of these transport equations are given in Fig.~\ref{fig:uubdgt}-\ref{fig:wwbdgt}. Among the turbulent energy spectra, only the streamwise component $E^z_{uu}$ has energy production from the mean flow, and the energy sources for the wall-normal and spanwise components $E^z_{vv}$ and $E^z_{ww}$ are the energy redistribution by the pressure-strain cospectra $\pi^z_{vv}$ and $\pi^z_{ww}$, respectively. Note here that 
\begin{align}
\pi^z_{uu} + \pi^z_{vv} + \pi^z_{ww} = 0
\end{align}
at any $y$-location and any wavelength $\lambda_z$. 

\begin{figure}[t]
    \centering
    \includegraphics[width=1\hsize]{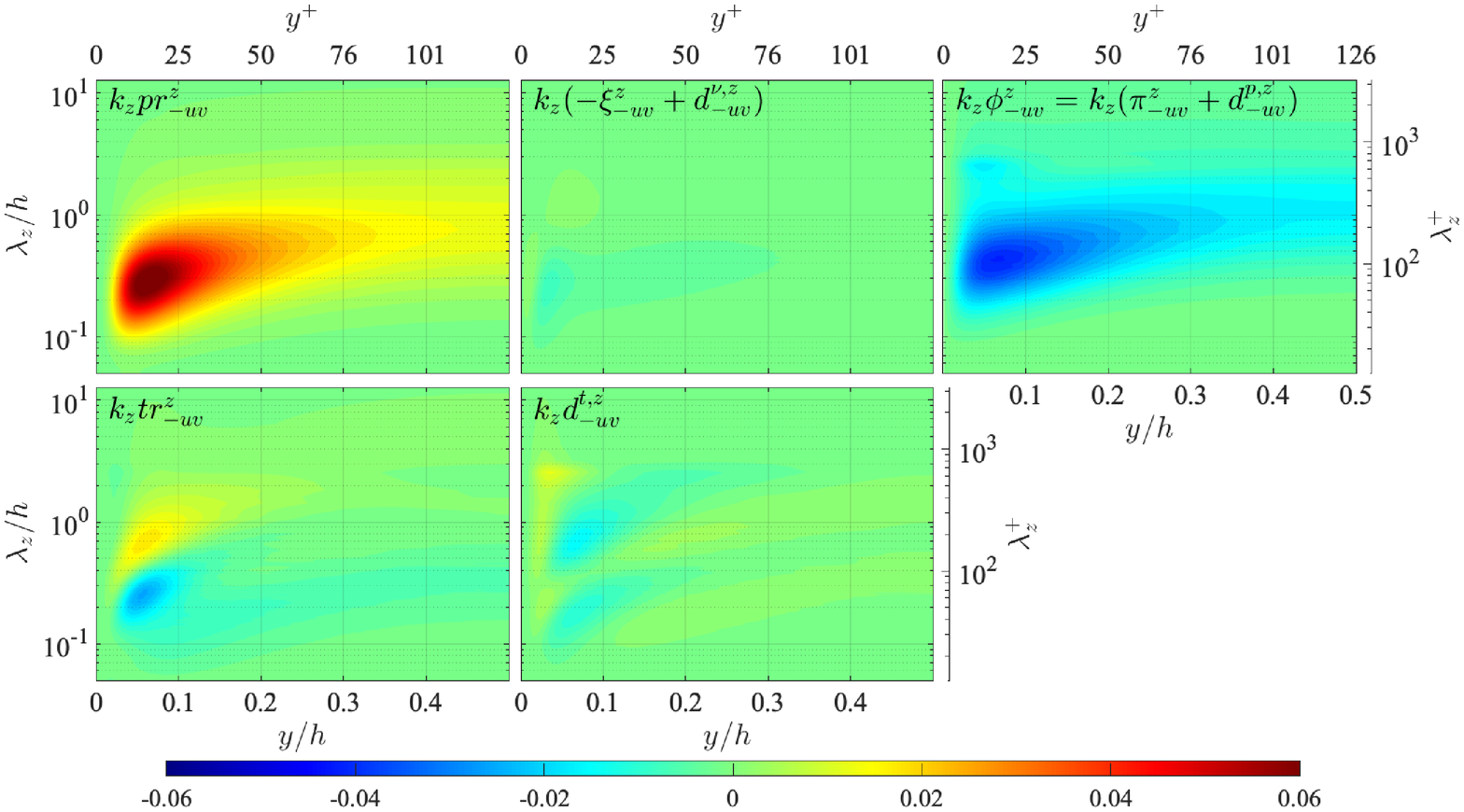}
    \caption{Distribution of terms in the transport equation of the Reynolds shear stress spectrum $E^z_{-uv}$, presented in the same manner as in Fig.~\ref{fig:utbdgt}: (top left)~production $k_z pr^z_{-uv}$; (top centre)~viscous terms $k_z(-\xi^z_{-uv}+d^{\nu,z}_{-uv})$; (top right)~pressure term $k_z \phi^z_{-uv}$; (bottom left)~interscale transport $tr^z_{-uv}$; (bottom centre)~turbulent diffusion $d^{t,z}_{-uv}$. The values are scaled by $u_\tau^4/\nu$.}
    \label{fig:uvbdgt}
\end{figure}

The transport equation of  the Reynolds shear stress spectrum is 
\begin{align}
\left( \pd{}{t}+U_k\pd{}{x_k} \right) E^z_{-uv} = pr^z_{-uv} - \xi^z_{-uv} + d^{\nu,z}_{-uv} + \phi^z_{-uv}  + d^{t,z}_{-uv} + tr^z_{-uv},
\end{align}
and the distributions of the budget terms are shown in Fig.~\ref{fig:uvbdgt}. 

\bibliography{library_kawata}

\end{document}